\documentclass[nomathfonts]{aipproc} 
\layoutstyle{6s}
\usepackage{xspace,graphicx,amsmath,amssymb,bbm}
\graphicspath{{Eps/}}
\title[Bayesian inference for inverse problems]{Bayesian inference for inverse problems}
\author{Ali Mohammad-Djafari}
{
  address = {Laboratoire des Signaux et Syst\`emes,\linebreak 
  Sup\'elec, Plateau de Moulon, 91192 Gif-sur-Yvette, France}
  ,email = djafari@lss.supelec.fr
}

\def\xs{\xspace}
\def\XS{\xspace}
\def\bm#1{\mbox{\boldmath $#1$}} 
\def\Ab{{\bm A}\XS}

\def\Gb{{\bm G}\XS}    
\def\Hb{{\bm H}\XS}    \def\hb{{\bm h}\XS}
\def\Ib{{\bm I}\XS}

    \def\mb{{\bm m}\XS}

\def\Pb{{\bm P}\XS}    
\def\Qb{{\bm Q}\XS}    
    \def\rb{{\bm r}\XS}
    \def\sb{{\bm s}\XS}

\def\Wb{{\bm W}\XS}    
\def\Xb{{\bm X}\XS}    \def\xb{{\bm x}\XS}
    \def\yb{{\bm y}\XS}
    \def\zb{{\bm z}\XS}

\def\epsilonb{{\bm \epsilon}\XS} 
			
\def\etab{{\bm \eta}\XS}			
\def\thetab{{\bm \theta}\XS}

\def\lambdab{{\bm \lambda}\XS}

\def\phib{{\bm \phi}\XS}

\def\Cc{\mbox{$\cal C$}\XS}	
\def\Dc{\mbox{$\cal D$}\XS}	
\def\Gc{\mbox{$\cal G$}\XS}	
\def\Lc{\mbox{$\cal L$}\XS}	
\def\Nc{\mbox{$\cal N$}\XS}	
\def\Sc{\mbox{$\cal S$}\XS}	
\def\babs{\begin{abstract}}             \def\eabs{\end{abstract}}
\def\barr{\begin{array}}                \def\earr{\end{array}}
\def\bcc{\begin{center}}                \def\ecc{\end{center}}

\def\bdes{\begin{description}}          \def\edes{\end{description}}
\def\bdoc{\begin{document}}             \def\edoc{\end{document}}
\def\ben{\begin{enumerate}}             \def\een{\end{enumerate}}
\def\beqn{\begin{eqnarray}}             \def\eeqn{\end{eqnarray}}
\def\beqnl#1{\beqn\label{#1}}           \def\eeqnl#1{\label{#1}\eeqn}
\def\beqnx{\begin{eqnarray*}}           \def\eeqnx{\end{eqnarray*}}
\def\bseqn{\begin{subeqnarray}}         \def\eseqn{\end{subeqnarray}}
\def\beq#1\eeq{\begin{equation}#1\end{equation}}
\def\beqx{$$}                           \def\eeqx{$$}
\def\bfig{\protect\begin{figure}}       \def\efig{\protect\end{figure}}
\def\bfigx{\protect\begin{figure*}}     \def\efigx{\protect\end{figure*}}
\def\bfl{\begin{flushleft}}             \def\efl{\end{flushleft}}
\def\bfr{\begin{flushright}}            \def\efr{\end{flushright}}
\def\bit{\begin{itemize}}               \def\eit{\end{itemize}}
\def\bpic{\begin{picture}}              \def\epic{\end{picture}}
\def\bqu{\begin{quote}}                 \def\equ{\end{quote}}
\def\bqun{\begin{quotation}}            \def\equn{\end{quotation}}
\def\bsl{\begin{slide}}                 \def\esl{\end{slide}}
\def\btabb{\begin{tabbing}}             \def\etabb{\end{tabbing}}
\def\btabl{\begin{table}}               \def\etabl{\end{table}}
\def\btablx{\begin{table*}}             \def\etablx{\end{table*}}
\def\btab{\begin{tabular}}              \def\etab{\end{tabular}}
\def\btabu{\begin{tabular}}             \def\etabu{\end{tabular}}
\def\btabx{\begin{tabular*}}            \def\etabx{\end{tabular*}}
\def\bbib{\begin{thebibliography}{999}} \def\ebib{\end{thebibliography}}
\def\bver{\begin{verbatim}}             \def\ever{\end{verbatim}}
\def\ie{\textit{i.e.,}\xs}
\def\eg{\textit{e.g.,}\xs}
\def\map{{\textit maximum a posteriori}\xs}
\def\apost{\textit{a posteriori}\xs}
\def\Apost{\textit{A posteriori}\xs}
\def\aprio{\textit{a priori}\xs}
\def\Aprio{\textit{A priori}\xs}
\def\etc{\textit{etc\dots}\xs}
\def\infine{\textit{in fine}\xs}
\def\adhoc{\textit{ad hoc}\xs}
\def\etal{\textit{et al.}\xs}
\def\pth#1{\left(#1\right)}
\def\acc#1{\left\{#1\right\}}
\def\cro#1{\left[#1\right]}
\def\bars#1{\left|#1\right|}
\def\norm#1{\left\|#1\right\|}
\def\diag{\mathop{\rm diag}}
\def\Diag{\mathop{\rm Diag}}
\def\tr{\mathop{\rm tr}}
\def\esp{\mathop{\rm E}}
\def\Esp#1{{\esp\cro{#1}}}
\def\argmax{\mathop{\rm arg\,max}}	
\def\argmin{\mathop{\rm arg\,min}}	
\def\xbh{\widehat{\xb}}
\def\zbh{\widehat{\zb}}
\def\ybh{\widehat{\yb}}
\def\sbh{\widehat{\sb}}
\def\xbt{\widetilde{\xb}}
\def\Pbh{\widehat{\Pb}}
\def\thetah{\widehat{\theta}}
\def\thetabh{\widehat{\thetab}}
\def\phibh{\widehat{\phib}}
\def\lambdabh{\widehat{\lambdab}}
\def\Xbh{\widehat{\Xb}}
\def\Hbh{\widehat{\Hb}}
\def\xh{\widehat{x}}
\def\hh{\widehat{h}}
\def\d#1{{\;\mbox{d}#1}}
\def\intg{\int\kern-1.1em\int}
\def\argmin#1#2{\mathop{\mbox{arg}\min}_{#1}\left\{{#2}\right\}}
\def\argmax#1#2{\mathop{\mbox{arg}\max}_{#1}\left\{{#2}\right\}}
\def\esp#1{\mbox{E}\left\{ #1 \right\}}
\def\espx#1#2{\mbox{E}_{#1}\left\{ #2 \right\}}
\def\dfdx#1#2{\frac{\mbox{d} #1}{\mbox{d} #2}}
\def\dpdx#1#2{\frac{\partial #1}{\partial #2}}
\def\expf#1{\exp\left[ {#1} \right]}
\def\disp{\displaystyle}
\def\lra{\longrightarrow}
\def\diag#1{\mbox{diag}\left[#1\right]}
\def\pmatrix#1{\left(\begin{matrix}#1\end{matrix}\right)}
\def\sumj{\sum_{j=1}^n}
\def\sumi{\sum_{i=1}^m}

\def\ms{mass spectrometry\xspace}
\def\Ms{Mass spectrometry\xspace}
\def\rem#1{}
\def\placefigure#1{\bigskip\centerline{\bf{#1}}\bigskip}

\def\defined{\,\shortstack{$\triangle$\\ =}\,}
\begin{document}
\begin{abstract}
Traditionally, the MaxEnt workshops start by a tutorial day. 
This paper summarizes my talk during 2001'th workshop at John Hopkins  
University. 
The main idea in this talk is to show how the Bayesian inference 
can naturally give us all the necessary tools we need to solve real 
inverse problems: starting by simple inversion where we assume to know 
exactly the forward model and all the input model parameters up to 
more realistic advanced problems of myopic or blind inversion where 
we may be uncertain about the forward model and we may have noisy data.  

Starting by an introduction to inverse problems through a few examples 
and explaining their ill posedness nature,  
I briefly presented the main classical deterministic 
methods such as data matching and classical regularization methods to 
show their limitations. I then presented the main classical probabilistic 
methods based on likelihood, information theory and maximum entropy 
and the Bayesian inference framework for such problems.    
I show that the Bayesian framework, not only generalizes all these 
methods, but also gives us natural tools, for example, 
for inferring the uncertainty of the computed solutions,  
for the estimation of the hyperparameters or for handling myopic 
or blind inversion problems. 
Finally, through a deconvolution problem example, 
I presented a few  state of the art methods based on Bayesian 
inference particularly designed for some of the 
mass spectrometry data processing problems. 
\end{abstract}
\maketitle
\section{Introduction}
\label{Introduction}
\subsection{Forward and inverse problems}
In experimental science, it is hard to find an example where we can 
measure directly a desired quantity. 
Describing \emph{mathematical models} 
to relate the measured quantities to the unknown quantity of interest 
is called \emph{forward modeling problem}. 
The main object of a forward modeling 
is to be able to generate data which are as likely as possible to the 
observed data if the unknown quantity was known. 
But, almost always, we want to use this model and the observed data to 
make inference on the unknown quantity of interest: 
This is the \emph{inversion problem}.  
To be more explicit, let take an example that we will use all along 
this paper to illustrate the different aspects of inverse problems. 
The example is taken from the \ms where the ideal physical 
quantity of interest is the components mass distribution of the 
material under the test. There are many techniques used in \ms. 
The Time-of-Flight (TOF) technique is one of them. 
In this technique, one measures the electrical 
current generated on the surface of a detector by the charged ions 
generated by the material under the test.  
Finding a very fine physical model to relate the time variation of this 
current to the distribution of the arrival times of the charged ions, 
which is itself 
related to the components mass distribution of the 
material under the test, is not an easy task. 
However, in a first approximation, assuming that the instrument is 
linear and its characteristics do not change during the acquisition time 
of the experiment, 
a very simple convolution model relates   
the raw data $g(t)$ to the unknown quantity of interest $f(t)$:  
\beq \label{1D_Convolution}
g(\tau)=\int f(t) \, h(\tau-t) \d{t},
\eeq
where $h(t)$ is the point spread function (psf) of the instrument. 
Figure~\ref{fig1} shows an example of data observed (signal in b) 
for a theoretical mass distribution (signal in a).    

\bfig[htb]
\btabu{@{}c@{}}
\includegraphics[width=\textwidth,height=4cm]{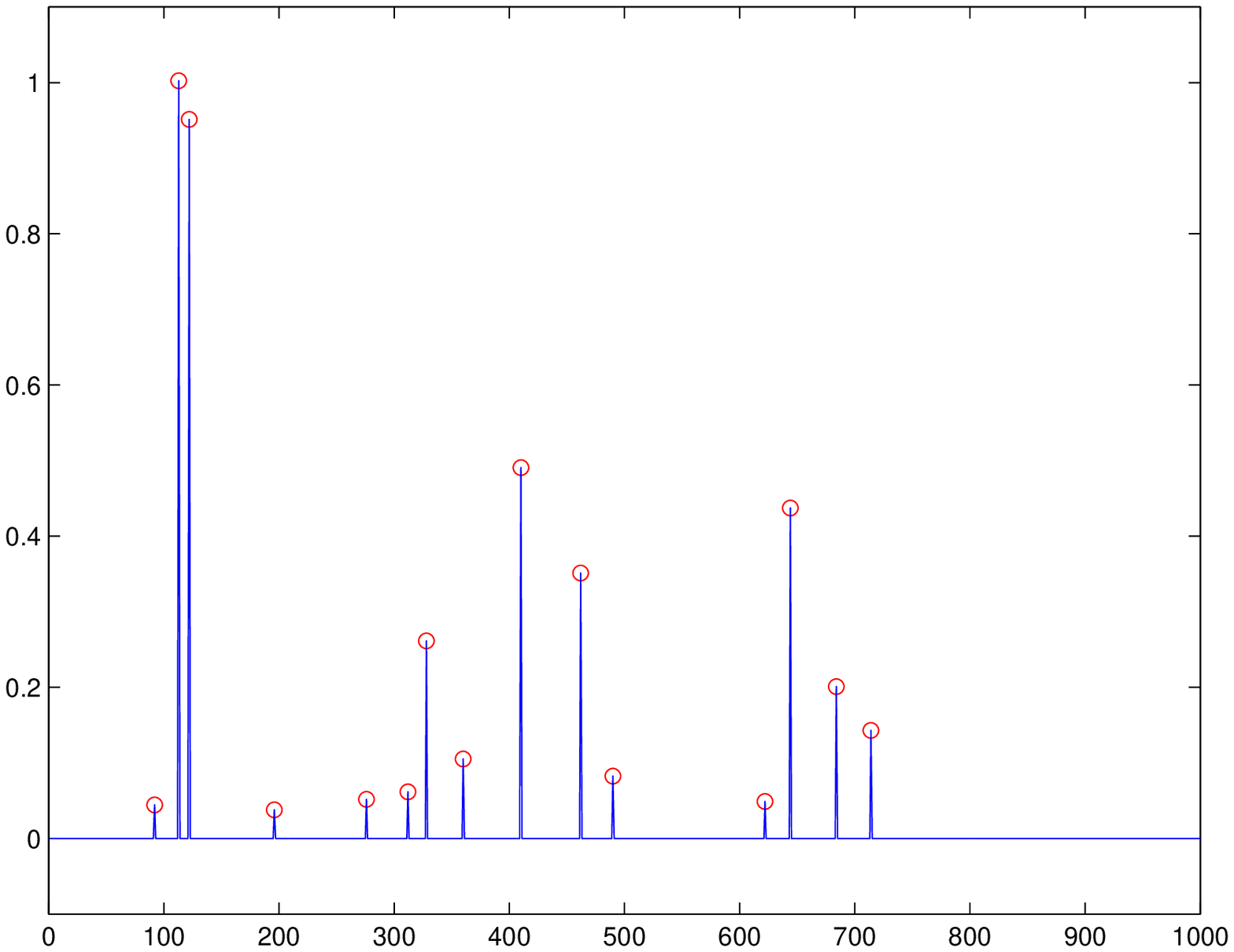}\\
\includegraphics[width=\textwidth,height=4cm]{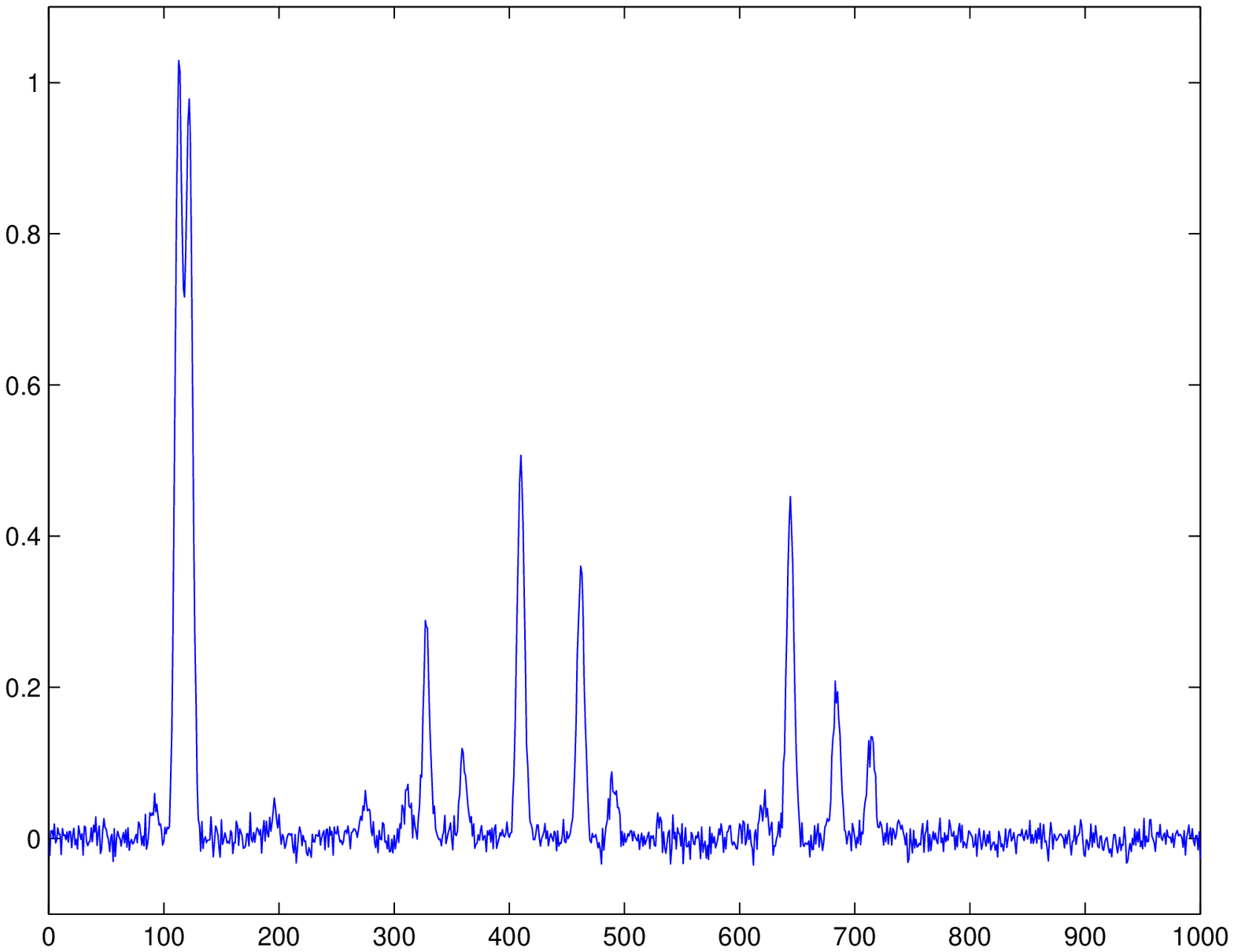}
\etabu
\caption{Blurring effect in TOF \ms data:~~ 
a) desired or theoretical spectrum,~~ 
b) observed data.}
\label{fig1}
\efig

In this example, the forward problem consists in computing $g$ 
given $f$ and $h$ which is given by a simple \emph{convolution} operation. 
The inverse problem of inferring $f$ given $g$ and $h$ is called 
\emph{deconvolution}, the inverse problem of inferring $h$ given $g$ 
and $f$ is called \emph{psf identification} and the inverse problem of 
inferring $h$ and $f$ given only $g$ is called 
\emph{blind deconvolution}. 

In my talk, I have given many more examples such as image restoration 
\beq \label{2D_Convolution}
g(x',y')=\int\int f(x,y) \, h(x'-x,y'-y) \d{x}\d{y}, 
\eeq
or Fourier synthesis inversion 
\beq \label{1D_FT}
g(\tau)=\int f(\omega) \, \exp\acc{-j\omega\tau} \d{\omega} 
\eeq
as well as a few non linear inverse problems. I am not going to detail 
them here, but I try to give a unified method to deal with all these 
problems. For this purpose, first we note that, in all these problems, 
we have always limited the number of data, for example 
$y_i=g(\tau_i), \; i=1,\ldots, m$. We also note that, to be able to do 
numerical computation, we need to model the unknown function $f$ by 
a finite number of parameters $\xb=[x_1,\ldots, x_n]$. 
As an example, we may assume that 
\beq \label{disc1}
f(t) = \sumj x_j b_j(t)  
\eeq
where $b_j(t)$ are known basis functions. 
With this assumption the raw data 
$\yb=[y_1,\ldots, y_m]$ are related to the unknown parameters $\xb$ by
\beq \label{disc2}
y_i=g(\tau_i)= \sumj H_{i,j} \, x_j 
\mbox{~~~with~~~} 
H_{i,j}=\intg b_j(t) h(t-\tau_i) \d{t}
\eeq
which can be written in the simple matrix form $\yb=\Hb \xb$. 
The inversion problem can then be simplified to the estimation of $\xb$ 
given $\Hb$ and $\yb$. 
Two approaches are then in competition: 
\\ 
i) the dimensional control approach which consists in an appropriate 
choice of the basis functions $b_j(\rb)$ and $n\le m$ in such a way that 
the equation $\yb=\Ab \xb$ be well conditioned; 
\\ 
ii) the more general regularization approach where a classical 
sampling basis for 
$b_j(\rb)$ with desired resolution is chosen no matter if $n>m$ or 
if $\Ab$ is ill conditioned. 
In the following, we follow the second approach which is more flexible 
for adding more general prior information on $\xb$. 

We must also remark that, in general, it is very difficult to give 
a very fine mathematical model to take account for all the different 
quantities affecting the measurement process. 
However, we can almost always  
come up with a more general relation such as 
\beq \label{disc3}
y_i=\hb_{\thetab}(\xb)+\epsilon_i, \quad i=1,\ldots, m 
\eeq
where $\thetab$ represents the unknown parameters of the forward model 
(for example the amplitude and the width of a Gaussian shape psf in a 
deconvolution problem) and $\epsilonb=[\epsilon_1,\ldots,\epsilon_m]$ 
represents all the errors (measurement noise, discretization errors and 
all the other uncertainties of the model). 
For the case of linear models we have 
\beq \label{model1}
\yb = \Hb_{\thetab} \xb + \epsilonb.
\eeq

In this paper we focus on this general problem. 
We first consider the case where the model is assumed to be perfectly 
known. This is the simple \emph{inversion problem}. 
Then we consider the more general case  
where we have also to infer on ${\thetab}$. 
This is the \emph{myopic} or \emph{blind inversion} problem. 

Even in the simplest case of perfectly known linear system and exact data: \\ 
i) the operator $\Hb$ may not be invertible ($\Hb^{-1}$ does not exist); 
\\  
ii) it may admit more than one inverse 
($\exists \Gb_1 \mbox{~and~} \Gb_2 | \Gb_1(\Hb)=\Gb_2(\Hb)=\Ib$ where $\Ib$ 
is the identity operator); or 
\\ 
iii) it may be very ill-posed or ill-conditioned (meaning that there exists 
$\xb$ and $\xb+\alpha\delta\xb$ \quad for which \quad  
$\norm{\Hb^{-1}(\xb)-\Hb^{-1}(\xb+\alpha\delta\xb)}$ \quad 
never vanishes even if $\alpha\mapsto 0$ 
\cite{Bertero88a,Demoment89}. 

These are the three necessary conditions of {\em existence}, 
{\em uniqueness} and {\em stability} of Hadamard for the well-posedness 
of an inversion problem. 
This explains the fact that, in general, even in this simple case, many na\"{\i}ve methods 
based on generalized inversion or on least squares may not give 
satisfactory results. 
The following figure shows, in a simple way, the 
ill-posedness of a deconvolution problem. 
On this figure, we see that three different input signals can result 
three outputs which are practically indistinguishable from each other. 
This means that, data matching alone can not distinguish between 
any of these inputs.

\bfig[htb]
\btabu{@{}c@{}}
\includegraphics[width=5.6cm,height=2.8cm]{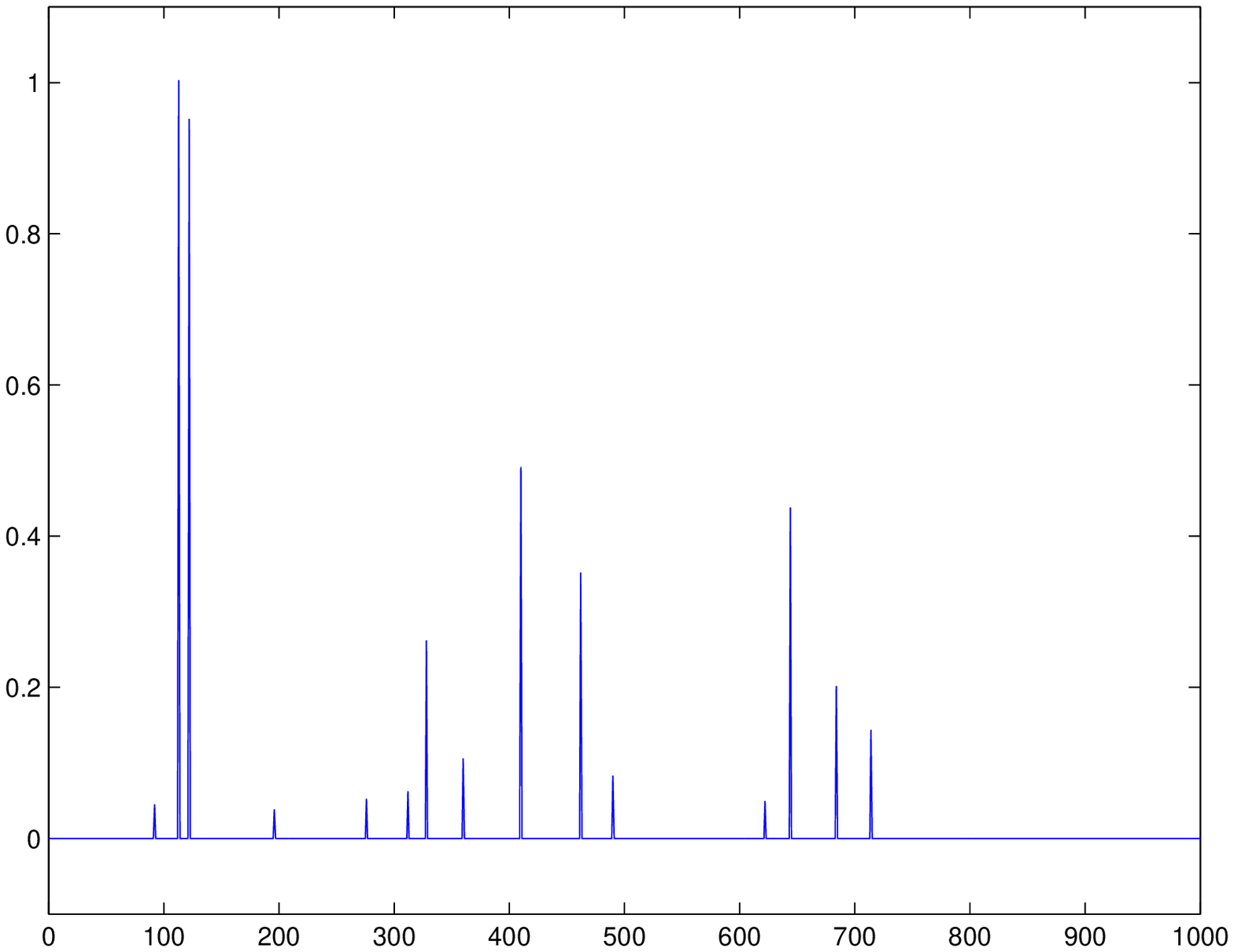}\\
\includegraphics[width=5.6cm,height=2.8cm]{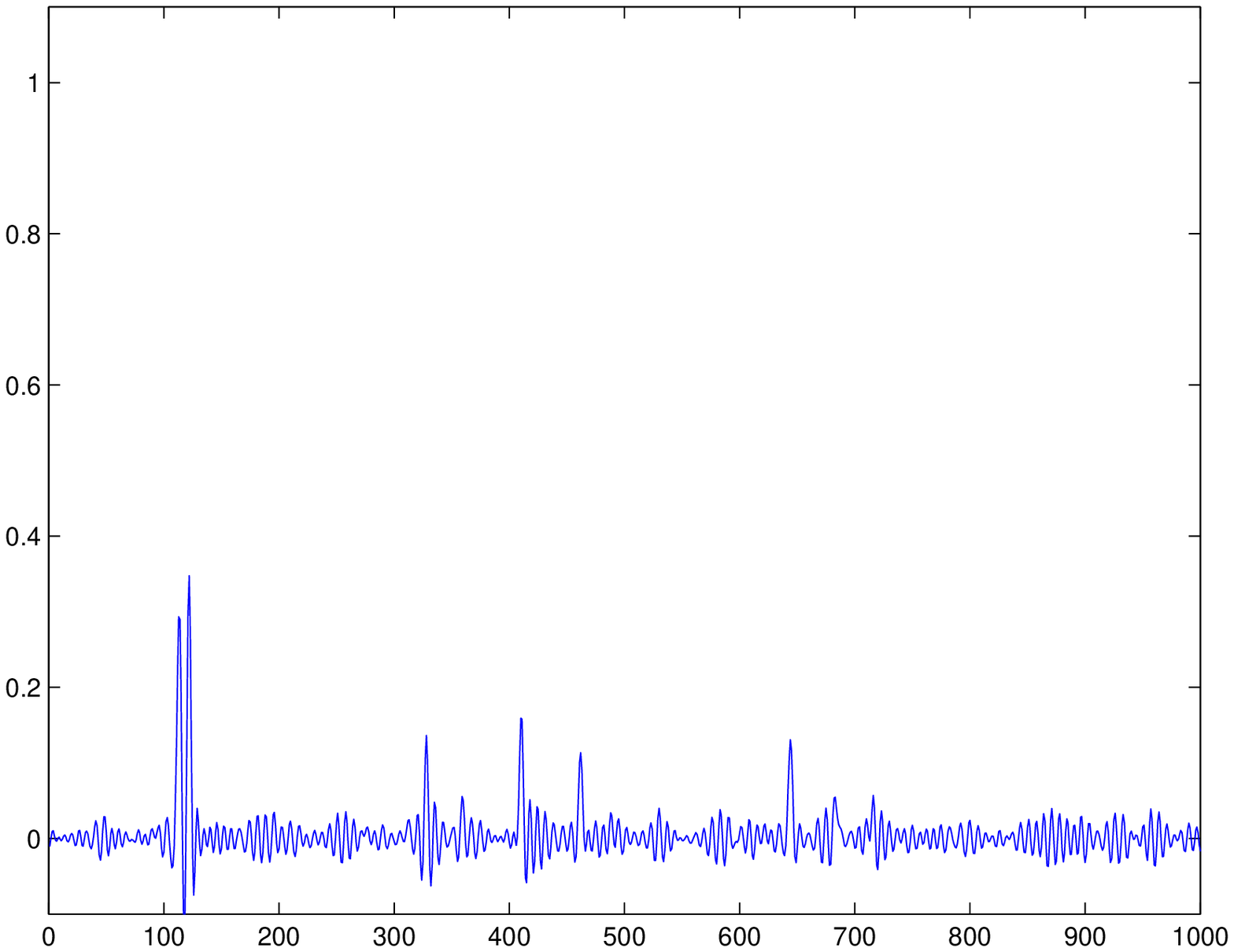}\\
\includegraphics[width=5.6cm,height=2.8cm]{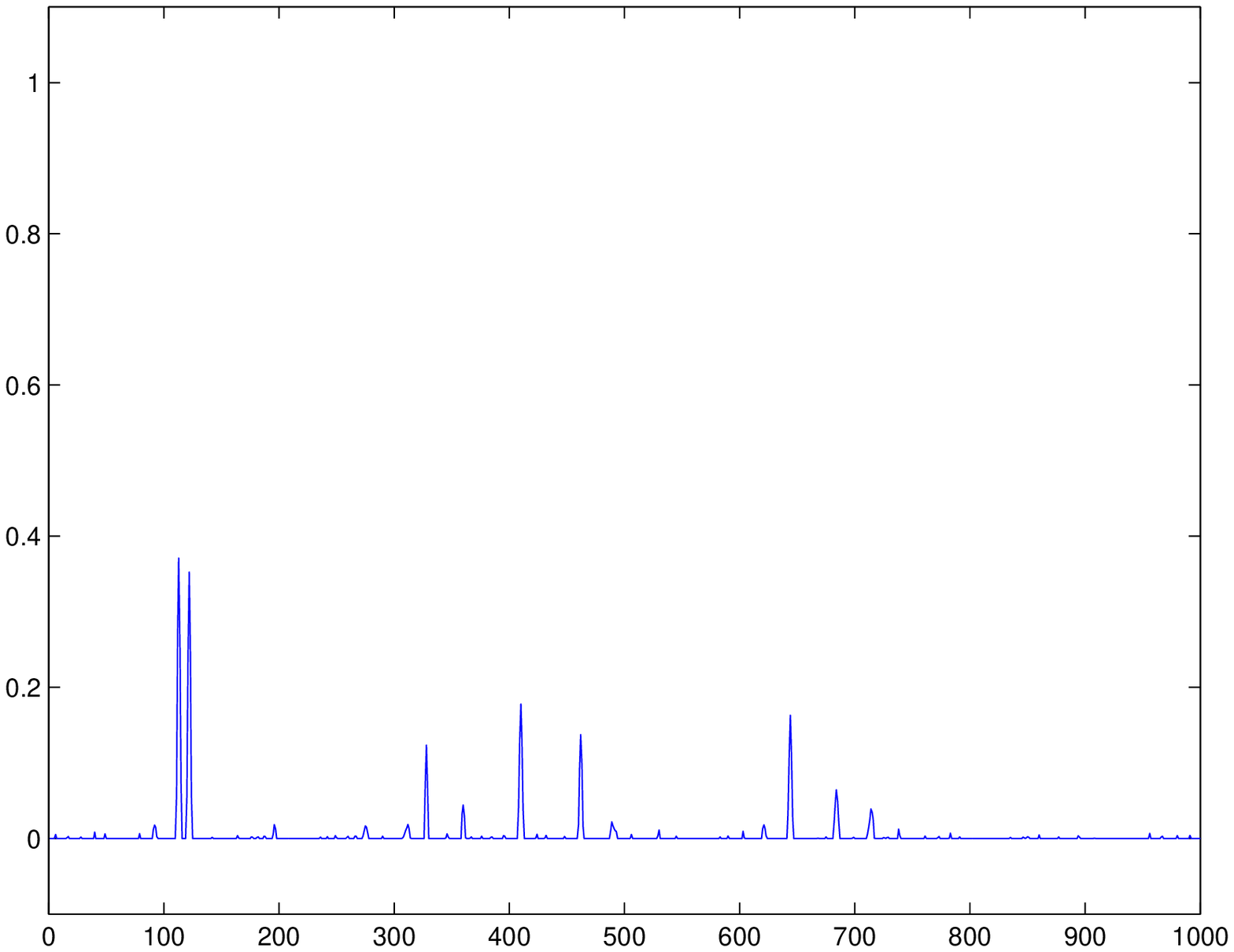}
\etabu
$\lra$
\btabu{@{}c@{}}
\includegraphics[width=2cm,height=2.8cm]{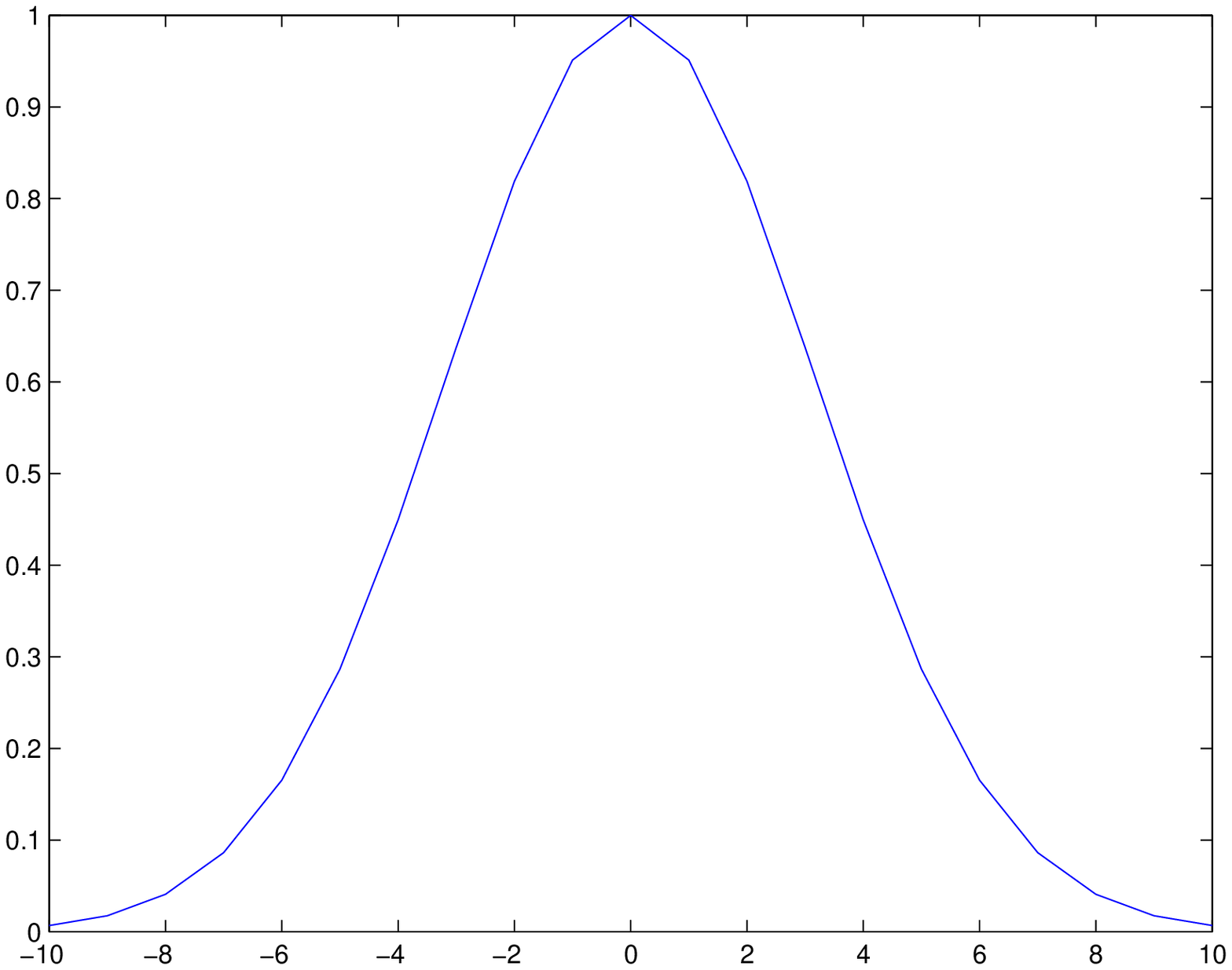}
\etabu
$\lra$
\btabu{@{}c@{}}
\includegraphics[width=5.6cm,height=2.8cm]{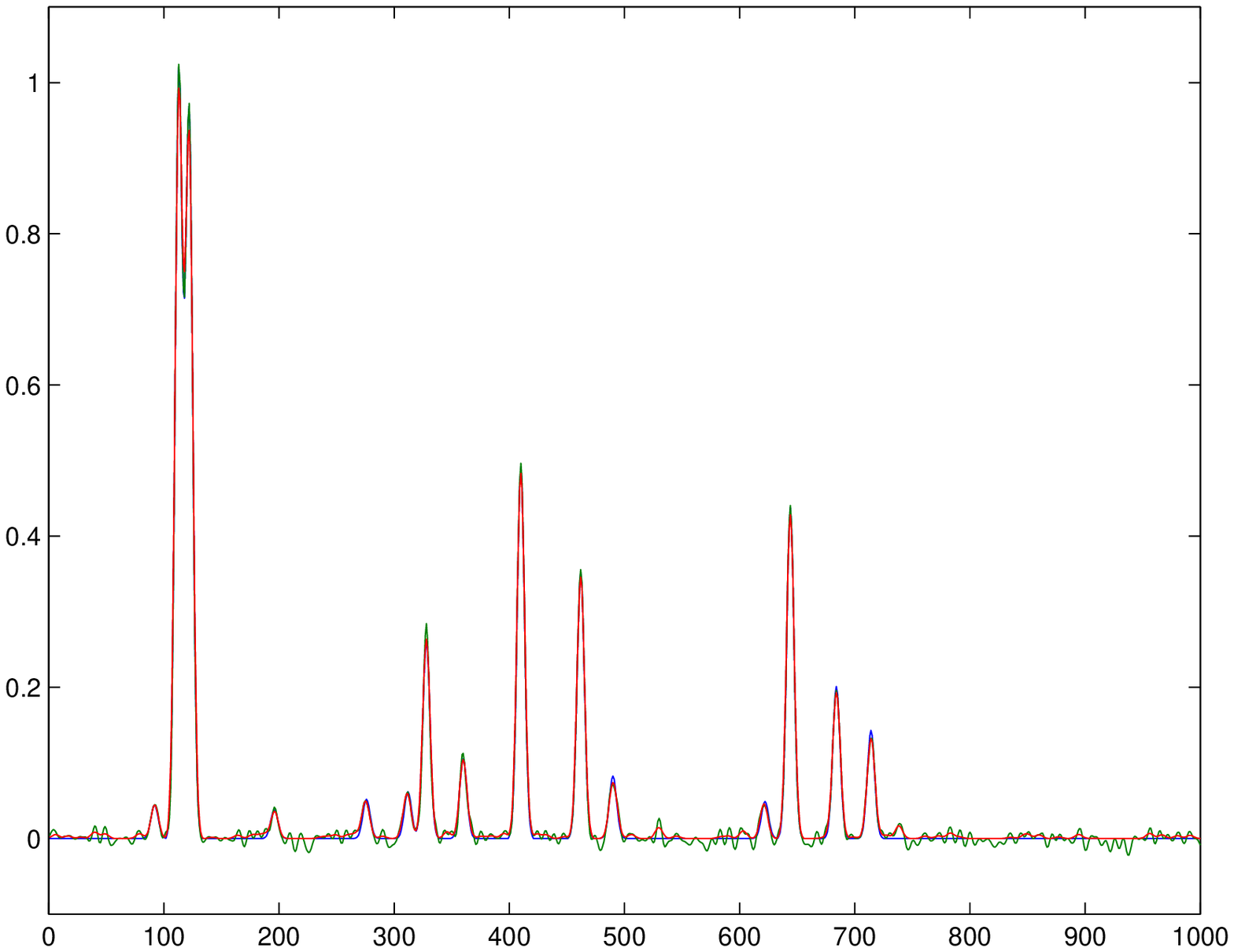}
\etabu
\caption{
Ill-posedness of a deconvolution problem: 
Inputs on the left give practically indistinguishable outputs.}
\label{figd}
\efig

As a conclusion, we see that, apart from the data, we need 
extra information. The art of \emph{inversion} in a particular inverse 
problem is how to include \emph{just enough prior information} to obtain 
a satisfactory result. In the following, first we summarize the 
classical deterministic approaches of data matching and regularization. 
Then, we focus on probabilistic approaches where errors and uncertainties 
are taken into account through the probability laws. 
Here, we distinguish, three classes of methods: those which only account 
for the data errors 
(error probability distribution matching and likelihood based methods), 
those which only account for uncertainties of unknown parameters (entropy 
based methods)  
and those which account for both of them (Bayesian inference approach).  

\section{Data matching and regularization methods}
\label{Regularization}

\subsection{Exact data matching} 
Let consider the discretized equation 
$y_i=h_i(\xb)+\epsilon_i, \; i=1,\ldots, m$; and 
assume first that the model and data are exact ($\epsilon_i=0$). 
We can then write $\yb=\hb(\xb)$.   

Assume now the system of equations is under determined, \ie there is more than one solution satisfying it (for example when the number of data is less than 
the number of unknowns). 
Then, one way to obtain a unique solution is to define an \aprio 
criterion, for example $\Delta(\xb,\mb)$ to choose that unique solution 
by 
\beq \label{Opt1}
\xbh=\argmin{\hb(\xb)=\yb}{\Delta(\xb,\mb)}
\eeq
where $\mb$ is an \aprio solution and $\Delta$ a distance measure. 

In the linear inverse problems case, 
the solution to this constrained optimization can be obtained via 
Lagrangian techniques which consists in defining the Lagrangian 
$\Lc(\xb,\lambdab)=\Delta(\xb,\mb)+\lambdab^t (\yb-\Hb\xb)$ 
and searching for $(\lambdabh,\xbh)$ through  
\beq
\left\{\barr{lcl} 
\lambdabh&=&\argmin{\lambdab}{\Dc(\lambdab)=\inf_{\xb}\Lc(\xb,\lambdab)}\\ 
\xbh     &=&\argmin{\xb}{\Lc(\xb,\lambdabh)}
\earr\right.
\eeq
Noting that 
$\nabla_{\xb} \Lc=\nabla_{\xb} \Delta(\xb,\mb) -\Hb^t\lambdab$ 
and 
$\nabla_{\lambdab} \Lc=\yb-\Hb\xb$ and defining 
$\Gc(\sb,\mb)=\sup_{\xb}\{\xb^t\sb-\Delta(\xb,\mb)\}$ 
the algorithm to find the solution $\xbh$ becomes: 
\\ 
-- Determine $\Gc(\sb,\mb)=\sup_{\xb}\{\xb^t\sb-\Delta(\xb,\mb)\}$;
\\ 
-- Find 
$\lambdabh
=\argmin{\lambdab}{\Dc(\lambdab)=\lambdab^t\yb-\Gc(\Hb^t\lambdab,\mb)}
$; 
\\ 
-- Determine $\xbh=\nabla_{\sb}\Gc(\Hb^t\lambdabh)$. 

As an example, when $\Delta(\xb,\mb)=\frac{1}{2}\norm{\xb-\mb}^2$ then 
$\Gc(\sb,\mb)=\mb^t\sb+\frac{1}{2}\norm{\sb}^2$, 
$\nabla_{\sb}\Gc=\mb+\sb$ 
and  
$\Dc(\lambdab)
=\lambdab^t\yb-\mb^t\Hb^t\lambdab+\frac{1}{2}\norm{\Hb^t\lambdab}^2
$ which results to   
$\lambdabh=(\Hb\Hb^t)^{-1} (\yb-\Hb\mb)$ 
and the solution is given by
\beq \label{Opt3}
\xbh=\mb + \Hb^t (\Hb\Hb^t)^{-1} (\yb-\Hb\mb).
\eeq
One can remark that, when $\mb=\bm{0}$ we have 
$\xbh=\Hb^t (\Hb\Hb^t)^{-1}\yb$ and this is the classical 
minimum norm generalized inverse solution. 

Another example is the classical \emph{Maximum Entropy} method case where $\Delta(\xb,\mb)=\mbox{KL}(\xb,\mb)$ 
is the Kullback-Leibler distance or cross entropy between $\xb$ and 
the \aprio solution $\mb$:  
\beq \label{cross-entropy}
\mbox{KL}(\xb,\mb)= \sum_j x_j\ln \frac{x_j}{m_j}-(x_j-m_j)
\eeq
Here, the solution is given by
\beq \label{Sol_MEC}
\xh_j=m_j \expf{-[\Ab^t\lambdabh]_j}
\mbox{~with~}
\lambdabh=\argmin{\lambdab}{\Dc(\lambdab)
=\lambdab^t\yb-\Gc(\Ab^t\lambdab,\mb)}
\eeq
where $\Gc(\sb,\mb)=\sum_j m_j\left(1 -\expf{-s_j}\right)$. 
But, unfortunately here $\Dc(\lambdab)$ is not a quadratic 
function of $\lambdab$ 
and thus there is not an analytic expression for $\lambdabh$. 
However, it can be computed numerically and many algorithms have 
been proposed for its efficient computation.  
See for example \cite{Skilling84} and the cited references 
for more discussions on the computational issues and algorithm 
implementation.  

The main issue here is that, this approach gives a satisfactory solution 
to the uniqueness of the inverse problem, but in general, the performances 
obtained by the resulting algorithms stay sensitive to error on the data. 

\subsection{Least squares data matching and regularization}
\label{REG2} 

When the discretized equation $\yb=\hb(\xb)$ is over-determined, 
\ie there is no solution satisfying it exactly (for example when the 
number of data is greater than the number of unknowns or when the data 
are not exact), one can try to estimate them by: 
\beq \label{LS-crit}
\xbh=\argmin{\xb}{\Delta(\yb, \hb(\xb))},
\eeq
where $\Delta(\yb, \hb(\xb))$ is a distance measure in the data space.  
The case where $\Delta(\yb, \hb(\xb))=\norm{\yb-\hb(\xb)}^2$ is 
the classical Least Squares (LS) criterion. 

For a linear inversion problem $\yb=\Hb \xb$, it is easy to see that 
any $\xbh$ which satisfies the normal equation $\Hb^t\Hb \xbh=\Hb^t\yb$ 
is a LS solution. 
If $\Hb^t\Hb$ is invertible and well-conditioned then 
$\xbh=(\Hb^t\Hb)^{-1}\Hb^t\yb$ is again the unique generalized 
inverse solution. 
But, in general, this is not the case: $\Hb^t\Hb$ is rank deficient 
and we need to constrain the space of the admissible solutions. 
The constraint LS is then defined as 
\beq \label{CLS-crit}
\xbh=\argmin{\xb\in\Cc}{\norm{\yb-\Hb\xb}^{2}}.
\eeq
where $\Cc$ is a convex set. The choice of the set $\Cc$ is primordial 
to satisfy the three conditions of a well-posed solution. 
An example is the positivity constraint: 
$\Cc=\{\xb : \; \forall j,\; x_j>0\}$. 
Another example is $\Cc=\{\xb : \; \norm{\xb}<\alpha\}$ where  
the solution can be computed via the optimization of 
\beq \label{CLS-equiv-crit}
J(\xb)=\norm{\yb-\Hb(\xb)}^{2}+\lambda \norm{\xb}.
\eeq
The main technical difficulty is the relation between $\alpha$ and 
$\lambda$. The minimum norm LS solution can also be computed using 
the singular value decomposition \cite{Hanson71}. 
The main issue here is that, even if this approach has been 
well understood and commonly used, it assumes implicitly that the noise 
and the $\xb$ are Gaussian. This may not be suitable in some 
applications, and more specifically in \ms data processing where the 
unknowns are spiky spectra. 

A more general regularization procedure is to define the 
solution to the inversion problem 
$\yb=\Hb(\xb)+\epsilonb$ as the optimizer of a compound criterion 
$J(\xb)=\norm{\yb-\Hb\xb}^2+\lambda\phi(\xb)$ or the more general 
criterion 
\beq \label{regularization}
J(\xb)=\Delta_1(\yb,\Hb\xb)+\lambda\Delta_2(\xb,\mb).
\eeq
where $\Delta_1$ and $\Delta_2$ are two distances or 
discrepancy measures, 
$\lambda$ a regularization parameter and $\mb$ an 
\aprio solution\cite{Idier96a}. 
The main questions here are: 
i) how to choose $\Delta_1$ and $\Delta_2$ and 
ii) how to determine $\lambda$ and $\mb$. 

For the first question, many choices exist:\\  
-- Quadratic or $L_2$ distance:\quad 
   $\Delta(\xb,\zb)=\norm{\xb-\zb}^2=\sum_j (x_j-z_j)^2$; 
\\ 
-- $L_p$ distance:\quad  
   $\Delta(\xb,\zb)=\norm{\xb-\zb}^p=\sum_j |x_j-z_j|^p$; 
\\ 
-- Kullback distance:\quad  
   $\Delta(\xb,\zb)=\sum_j x_j \ln (x_j/z_j) - (x_j-z_j)$; 
\\ 
-- Roughness distance:\quad $\Delta(\xb,\zb)$ any of the previous 
distances with $z_j=x_{j-1}$ or $z_j=(x_{j-1}+x_{j+1})/2$ or any 
linear function  
$z_j=\psi(x_k, k\in \Nc(j))$ where $\Nc(j)$ stands for 
the neighborhood of $j$.   
(One can see the link between this last case and the Gibbsian energies 
in the Markovian modeling of signals and images.) 

The second difficulty in this deterministic approach is the 
determination of the regularization parameter $\lambda$. 
Even if there are some techniques based on cross validation 
\cite{Titterington85,Golub79,Fortier93}, there is not natural tools for 
their extension to other hyperparameters in a natural way. 

As a simple example, we consider the case where both $\Delta_1$ 
and $\Delta_2$ are quadratic: 
$J(\xb)=\norm{\yb-\Hb\xb}_{\Wb}^2+\lambda\norm{\xb-\mb}_{\Qb}^2$. 
The optimization problem, in this case, has an analytic solution: 
\beq
\xbh=\pth{\Hb^t\Wb\Hb+\lambda\Qb}^{-1}\pth{\Hb^t\Wb\yb-\Qb\mb)}
\eeq 
which can also be written 
\beq
\xbh=\mb+\Qb^{-1}\Hb^t\pth{\Hb\Qb^{-1}\Hb^t+\lambda^{-1}\Wb^{-1}}^{-1}
         \pth{\yb-\Hb\mb}
\eeq 
which is a linear function of the \aprio solution $\mb$ and the 
data $\yb$. Note also that when $\mb=\bm{0}$, $\Qb=\Ib$ and $\Wb=\Ib$ 
we have $\xbh=\pth{\Hb^t\Hb+\lambda\Ib}^{-1}\Hb^t\yb$ or 
$\xbh=\Hb^t\pth{\Hb\Hb^t+\lambda^{-1}\Ib}^{-1}\yb$ and when $\lambda=0$ 
we obtain the generalized inverse solutions 
$\xbh=\pth{\Hb^t\Hb}^{-1}\Hb^t\yb$ or  
$\xbh=\Hb^t\pth{\Hb\Hb^t}^{-1}\yb$. 

As we mentioned before, the main practical difficulties in this 
approach are the choice of $\Delta_1$ and $\Delta_2$ and the determination 
of the hyperparameters $\lambda$ and the inverse covariance matrices $\Wb$ 
and $\Qb$. 

As a main conclusion on these deterministic inversion methods, we can say 
that, even if, in practice, they are used and give satisfaction, they lack 
tools to handle with uncertainties and to account for more precise \aprio knowledge of statistical properties of errors and unknown parameters. 
The probabilistic methods can exactly handle more easily these problems 
as we will see in the following. 

\section{Probabilistic methods}
\label{Probabilistic_methods}

\subsection{Probability distribution matching and maximum likelihood} 
\label{Maximum_likelihood}

The main idea here is to account for data and model uncertainty through 
the assignment of a theoretical distribution $p_{Y|X}(\yb|\xb)$ to the data. 
In probability distribution matching method, the main idea is to determine 
the unknown parameters $\xb$ by minimizing a distance measure 
$\Delta(\rho, p)$ between the empirical histogram $\rho$ of the data 
defined as 
\beq
\rho(\zb) \defined \frac{1}{N} \sum_i \delta (z_i-y_i) 
\eeq
and the theoretical distribution of the data $p_{Y|X}(\zb|\xb)$. 

When $\Delta(p,\rho)$ is choosed to be the Kullback-Leibler mismatch  
measure 
\beqn
KL[\rho,p] 
&\defined& 
\intg \rho(\zb) \ln \frac{\rho(\zb)}{p_{Y|X}(\zb|\xb)} \d{\zb} \nonumber \\ 
&=& -\intg \rho(\zb) \ln p_{Y|X}(\zb|\xb) \d{\zb} 
    + \intg \rho(\zb) \ln \rho(\zb) \d{\zb} 
\eeqn
we have 
\beq
\xbh=\argmin{\xb}{KL\left[\rho,p\right]} 
=\argmin{\xb}{-\intg \rho(\zb) \ln p_{Y|X}(\zb|\xb) \d{\zb}}.
\eeq
It is then easy to see that, for the i.i.d. data, this estimate becomes 
equivalent to the maximum likelihood (ML) estimate 
\beq
\xbh=\argmin{\xb}{-\ln p_{Y|X}(\zb|\xb)|_{\zb=\yb}}=\argmax{\xb}{p_{Y|X}(\yb|\xb)}.
\eeq
In the case of a linear model and Gaussian noise, it is easy to show 
that the ML estimate becomes equivalent to the LS one, which in general, 
does not give satisfactory results as we have discussed it in the previous 
section.  

The important point to note here is that, in this approach, only the 
data uncertainty is considered and modeled through the probabilty law 
$p_{Y|X}(\yb|\xb)$. 
We will see in the following that, in contrary to this approach, 
in information theory and maximum entropy methods, the data 
and model are assumed to be exact and only the uncertainty of $x$ 
is modeled through an \aprio reference measure $\mu(\xb)$ which is 
updated to an \apost probabilty law $p(\xb)$ by optimizing the KL 
mismatch $\mbox{KL}(p,\mu)$ subject to the data constraints. 

\subsection{Maximum entropy in the mean}
The main idea in this approach is to consider $\xb$ as the mean value 
of a quantity $\Xb\in\Cc$, 
where $\Cc$ is a compact set on which we want to define a probability 
law $P$: $\xb=\espx{P}{\Xb}$  
and the data $\yb$ as exact equality constraints on it:
\beq 
\yb=\Hb \xb=\Hb \espx{P}{\Xb}=\intg_{\Cc} \Hb \xb \d{P(\xb)}. 
\eeq 
Then, assuming that we can translate our prior information on 
the unknowns  through a prior law (a reference measure) $\d{\mu(\xb)}$,  
we can determine the distribution $P$ by: 
\beq 
\hbox{maximize}\quad  -\intg_{\Cc} \ln  
\frac{\d{P(\xb)}}{\d{\mu(\xb)}} \d{P(\xb)} 
\quad\hbox{s.t.}\quad  \yb=\Hb \xb=\Hb \espx{P}{\Xb}. 
\eeq
The solution is obtained via the Lagrangian: 
\[
\disp{
\Lc(\xb,\lambdab) 
= \intg_{\Cc} \left[ \ln \frac{\d{P(\xb)}}{\d{\mu(\xb)}} 
- \lambdab^t (\yb-\Hb \xb) \right] \d{P(\xb)} 
}
\] 
and is given by:\quad 
\(
\d{P(\xb,\lambdab)}=\expf{\lambdab^t [\Hb \xb]-\ln  
Z(\lambdab)}\d{\mu(\xb)}, 
\) \quad where
\\ 
\(\disp{
 Z(\lambdab)=\intg_{\Cc} \expf{\lambdab^t [\Hb \xb]} \d{\mu(\xb)} 
}\). 
The Lagrange parameters are obtained by searching the unique 
solution (if exists) of the following system of non linear equations:
\beq
 \dpdx{\ln Z(\lambdab)}{\lambda_i}=y_i,\quad i=1,\cdots,M.
\eeq
Then, naturally, the solution to the inverse problem is defined 
as the expected value of this distribution: \quad 
\(
 \xbh(\lambdab)=\espx{P}{\Xb}=\int \xb \, \d{P(\xb,\lambdab)}.  
\) 
The interesting point here is that, the solution 
$\xbh(\lambdabh)$ can be computed without actually computing $P$ 
in two ways: 

\noindent 
-- Via optimization of a dual criterion: The solution $\xbh$ is expressed 
as a function of the dual variable $\sbh=\Hb^t\lambdabh$ by 
\(
\xbh(\sbh)=\nabla_{\sb}{G(\sbh,\mb)}
\) 
where 
\[
G(\sb,\mb)=\ln Z(\sb,\mb)=\ln \intg_{\Cc} \expf{\sb^t \xb} \d{\mu(\xb)}, 
\quad 
\mb=\espx{\mu}{\Xb}=\intg_{\Cc} \xb \d{\mu(\xb)}
\] 
and \qquad 
\(\disp{
\lambdabh=\argmax{\lambdab}{D(\lambdab)=\lambdab^t\yb-G(\Hb^t\lambdab)}
}\). 
\\ 
-- Via optimization of a primal or direct criterion: 
\[
\xbh=\argmin{\xb\in{\Cc}}{H(\xb,\mb)} \hbox{~~s.t.~~}\yb=\Hb\xb 
\mbox{~~where~~} 
H(\xb,\mb)=\sup_{\sb}\{\sb^t \xb - G(\sb,\mb)\}.
\]
Another interesting point is the link between these two options:  
\\ 
i) Functions $G$ and $H$ depend on the reference measure $\mu(\xb)$;  
\\ 
ii) The dual criterion $D(\lambdab)$ depends on the data and the 
function $G$;
\\ 
iii) The primal criterion $H(\xb,\mb)$ is a distance measure between 
$\xb$ and $\mb$ which means: 
$H(\xb,\mb) \ge 0$ and $H(\xb,\mb)=0 \quad \hbox{iff}\quad \xb=\mb$;
\quad   
$H(\xb,\mb)$ is differentiable and convex on $\Cc$ and   
$H(\xb,\mb)=\infty$ \quad if $\xb\not\in {\Cc}$;  
\\ 
iv) If the reference measure is separable: 
$\mu(\xb)=\prod_{j=1}^N \mu_j(x_j)$
then $P$ is too: \\ 
$\d{P(\xb,\lambdab)}=\prod_{j=1}^N \d{P_j(x_j,\lambdab)}$ 
and we have
\[
G(\sb,\mb)=\sum_j g_j\left(s_j,m_j\right), \quad 
H(\xb,\mb)=\sum_j h_j(x_j,m_j), \quad 
\xh_j=g'_j(s_j,m_j). 
\]
where $g_j$ is the log Laplace transform (Cramer transform) of $\mu_j$: 
\[
g_j(s)=\ln \int \expf{s x} \d{\mu_j(x)}; 
\] 
and $h_j$ is the convex conjugate of $g_j$: \quad 
   $\disp{h_j(x)=\max_{\sb}\{s x -g_j(s)\}}$.

The following table gives three examples of choices for $\mu_j$ 
and the resulting expressions for $g_j$ and $h_j$:  
\[
\barr[width=\textwidth]{l|c|c|c}
& \mu_j(x) & g_j(s) & h_j(x,m) 
\\ \cline{1-4} \cline{1-4}
\hbox{Gaussian:} 
 & \disp{\expf{- \pth{{1}/{2}}(x-m)^2}}        
 & \disp{\pth{{1}/{2}}(s-m)^2}  
 & \disp{\pth{{1}/{2}}(x-m)^2}  
\\ \cline{1-4}
\hbox{Poisson:} 
 & \disp{\pth{{m^x}/{x!}}\expf{-m}} 
 & \disp{\expf{m-s}}
 & \disp{-x\ln\pth{{x}/{m}}+m-x} 
\\ \cline{1-4}
\hbox{Gamma: }\hspace*{5mm} 
 & \disp{ x^{\alpha-1} \expf{-\pth{{x}/{m}}}} 
 & \disp{\ln (s-m)} 
 & \disp{-\ln\pth{{x}/{m}}+\pth{{x}/{m}}-1}          
\\ \cline{1-4}\cline{1-4}
\earr
\]
We may remark that the two famous expressions of the Burg 
$\ln x $ and Shannon $- x \ln x$ 
entropies are obtained as special cases. 

As a conclusion, we see that the Maximum entropy in mean extends 
in some way the classical ME approach by giving other expressions 
for the criterion to optimize. Indeed, it can be shown that when  
we optimize a convex criterion subject to the data 
constraints we 
are optimizing the entropy of some quantity related to the unknowns 
and \emph{vise versa}. 
However, as we have mentioned, basically, in this 
approach the data and the model are assumed to be exact even if some 
extensions to the approach gives the possibility to account 
for the errors \cite{LeBesnerais99}. 
In the next section, we see how the Bayesian approach 
can naturally account for both uncertainties on the data and on 
the unknown parameters $\xb$. 

\section{Bayesian inference approach}
\label{Bayesian}
In Bayesian approach, the main idea is to translate our prior knowledge 
on the errors $\epsilonb$ and on the unknowns $\xb$ to prior probability 
laws $p(\epsilonb)$ and $p(\xb)$. The next step is to use the forward 
model and $p(\epsilonb)$ to deduce $p(\yb|\xb)$. 
The Bayes rule can then be used to determine the posterior law of 
the unknowns $p(\xb|\yb)$ 
from which we can deduce any information about the unknowns 
$\xb$. The posterior $p(\xb|\yb)$ is thus the final product of 
the Bayesian approach. However, very often, we need a last step 
which is to take out the necessary information about $\xb$ from this 
posterior. The tools for this last step are the 
decision and estimation theories.  

To illustrate this, let consider the case of linear inverse 
problems \(\yb=\Hb\xb+\epsilonb\). 
The first step is to write down explicitly our hypothesis: 
starting by the hypothesis that 
$\epsilonb$ is zero-mean (no systematic error), white (no time correlation 
for the errors) and assuming that we may only have some idea about its 
energy $\sigma_{\epsilon}^2=1/(2\phi_1)$, and using either the intuition or 
the Maximum Entropy Principle (MEP) lead to a Gaussian prior law: 
$\epsilonb\sim{\Nc}\left(\bm{0},1/(2\phi_1) \Ib\right)$. 
Then, using the forward model with this assumption leads to:
\beq
p(\yb|\xb,\phi_1)\propto \expf{-\phi_1 \|\yb-\Hb\xb\|^2}.
\eeq
The next step is to assign a prior law to the unknowns $\xb$. 
This step is more difficult and needs more caution. 
In inverse problems, as we presented, $\xb$ represents the samples 
of a signal or the pixel values of an aerian image. 
Very often then we have ensemblist prior knowledge about the  
signals or images concerned by the application and we can 
model them. The art of the engineer is then to choose the 
appropriate model and to translate this information to a 
probability law to reflect it. 
 
Again here, let illustrate this step, first through a few general 
examples and then more specifically the case of \ms deconvolution problem. 

In the first example, we assume that, \aprio we do not have 
(or we do not want or we are not able to account for) 
any knowledge about the correlation between the components of $\xb$. 
This leads us to 
\beq
p(\xb)=\prod_j p_j(x_j). 
\eeq
Now, we have to assign $p_j(x_j)$. 
For this, we may assume to know the mean 
values $m_j$ and some idea about the dispersions about these mean values. 
This again leads us to Gaussian laws $\Nc(m_j,\sigma_{x_j}^2)$, and 
if we assume the same dispersions $\sigma_{x_j}^2=1/(2\phi_2), \forall j$ 
we obtain 
\beq \label{G_Prior}
p(\xb)\propto \expf{-\phi_2 \sum_j \bars{x_j-m_j}^2} 
=\expf{-\phi_2 \norm{\xb-\mb}^2}
\eeq
With these assumptions, using the Bayes rule, we obtain 
\beq \label{G_Posterior}
p(\xb|\yb)\propto \expf{-\phi_1\norm{\yb-\Hb\xb}^2-\phi_2\norm{\xb-\mb}^2}. 
\eeq
This posterior law contains all the information we can have on $\xb$ 
(combination of our prior knowledge and data). 
If $\xb$ was a scalar or a vector of only two components, we could 
plot the probability distribution and look at it. 
But, in practical applications, 
$\xb$ may be a vector with huge number of components. 
Then, even if we can obtain an expression for this posterior, we 
may need to summarize its information content. 
In general then, we may choose, equivalently, between summarizing it 
by its mode, mean, marginal modes, \etc, or use the decision and estimation 
theory to define \emph{point estimators} to be used to compute   
(\emph{best representing values}). 
For example, we can choose the value $\xbh$ which corresponds to the 
mode of $p(\xb|\yb)$-- the \emph{Maximum  a posteriori (MAP)} estimate, 
or the value $\xbh$ which corresponds to the mean of this posterior-- 
the \emph{Posterior mean (PM)} estimate, or when interested to the 
component $x_j$, to choose $\xh_j$ corresponding to the mode of the 
posterior marginal $p(x_j|\yb)$. 

We can also generate samples from this posterior and just look at them 
as a movie or use them to compute the PM estimate. We can also use it 
to compute the posterior covariance matrix 
$(\Pb=\esp{(\xb-\xbh)(\xb-\xbh)^t}$ where $\xbh$ 
is the posterior mean), 
from which we can infer 
on the uncertainty of the proposed solutions. 

In the Gaussian priors case we just presented, it is easy to see that, 
the posterior law is also Gaussian and all these estimates are the same 
and can be computed by minimizing 
\beq \label{LoG_Posterior}
J(\xb)=-\ln p(\xb|\yb)=\norm{\yb-\Hb\xb}^2 + \lambda \norm{\xb-\mb}^2 
\quad\mbox{with~~}
\lambda=\frac{\phi_2}{\phi_1}=\frac{\sigma_{\epsilon}^2}{\sigma_{x}^2}. 
\eeq
We may note here the analogy with the quadratic regularization 
criterion (\ref{regularization}) with the emphasis that 
the choice $\Delta_1(\yb, \Hb\xb)=\norm{\yb-\Hb\xb}^2$ 
and $\Delta_2(\xb, \mb)=\norm{\xb-\mb}^2$ are the direct consequences 
of Gaussian choices for prior laws of the noise $p(\epsilonb)$ 
and the unknowns $p(\xb)$. 

The Gaussian choice for $p_j(x_j)$ is not always a pertinent one. 
For example, we may \aprio know that the distribution of $x_j$ around 
their means $m_j$ are more concentrated but great deviations from them 
are also more likely than a Gaussian distribution. This knowledge can be 
translated by choosing a Generalized Gaussian law: 
\beq \label{GG_Prior}
p(x_j)\propto \expf{-\frac{1}{2\sigma_{x}^2}\bars{x_j-m_j}^{p}}, \quad 
1\le p \le 2. 
\eeq

In some cases we may know more, for example we may know that $x_j$ are 
positive values. Then a Gamma prior law 
\beq
p(x_j)=\Gc(\alpha,m_j)\propto 
(x_j/m_j)^{-\alpha}\expf{-x_j/m_j} 
\eeq
would be a better choice. 

In some other cases we may know that $x_j$ are 
discrete positive values. Then a Poisson prior law 
\beq
p(x_j)\propto \frac{m_j^{x_j}}{x_j!} \expf{-m_j}
\eeq
is a better choice. 

In all these cases, the expression of the posterior is 
$p(\xb|\yb)\propto \expf{-J(\xb)}$ with  
$J(\xb)=\norm{\yb-\Hb\xb}^2 +\lambda \phi(\xb)$ 
where $\phi(\xb)=-\ln p(\xb)$. It is interesting to note the different 
expressions of $\phi(\xb)$ for the prior laws discussed and remark that 
they contain different \emph{entropy expressions} for the $\xb$. 

The last general example is the case where \aprio we know that $x_j$ 
are not independent, for example when they 
represents the pixels of an aerian image. 
We may then use a Markovian modeling 
\beq
p(x_j|x_k, k\in\Sc)=p(x_j|x_k, k\in\Nc(j)), 
\eeq
where $\Sc=\{1,\ldots, N\}$ stands for the whole set of pixels and 
$\Nc(j)=\{k : |k-j|\le r\}$ stands for $r$-th order neighborhood of $j$. 

With some assumptions on the border limits, such models again result 
to the optimization of the same criterion with 
\beq \label{Markovian_Prior}
\phi(\xb)=\Delta_2(\xb,\zb)=\sum_j \phi(x_j,z_j) 
\hbox{~where~} 
z_j=\psi(x_k, k\in \Nc(j))
\eeq
with different potential functions $\phi(x_j,z_j)$.  

A simple example is the case where $z_j=x_{j-1}$ and $\phi(x_j,z_j)$ 
any function in between the following:
\[
\left\{
|x_j-z_j|^{\alpha}, \quad 
\alpha \ln\frac{x_j}{z_j} + \frac{x_j}{z_j}, \quad 
x_j\ln \frac{x_j}{z_j}+(x_j-z_j)
\right\}
\] 
See (\cite{Djafari93b,Brette94,Brette94a,Brette94b}) 
for some more discussion and properties of 
these potential functions. 

As one of the main conclusions here, we see that, as it concerns the MAP 
estimation, the Bayesian approach is equivalent to the general 
regularization. However, here the choice of the distance measure 
$\Delta_1(\yb,\Hb\xb)$ depends on the forward model and the hypothesis 
on the noise and the choice of the distance measure 
$\Delta_2(\xb,\mb)$ depends on the prior law chosen for $\xb$. 

One more extra feature here is that, we have access to the whole posterior 
$p(\xb|\yb)$ from which, not only we can define an estimate but also, 
we can quantify its corresponding remained uncertainty. We can also 
compare posterior and prior laws of the unknowns to measure the amount 
of information contained in the observed data. Finally, as we will see 
in the following, we have finer tools to model unknown signals or images 
and to estimate the hyperparameters. 

\section{Open problems and advanced methods}
As we have remarked in previous sections, in general, the solution of 
an inverse problem depends on our prior hypothesis on errors $\epsilonb$ 
and on $\xb$. Before applying the Bayes rule, we have to assign the prior 
laws to them. From the forward model and assumptions on $\epsilonb$ we 
assign $p(\yb|\xb,\phib_1)$ and from the assumptions on $\xb$ we 
assign $p(\xb|\phib_2)$. 
This step is one of the most crucial part of the applicability of the 
Bayesian framework for inverse problems. Modeling a signal and finding the corresponding expression for the prior law $p(\xb|\phib_2)$ is not an 
easy task. This choice may have many consequences: the complexity of 
the computation of the posterior and consequently the computation of 
any point estimators such as MAP (which needs optimization) 
or PM (which needs integration either analytically or by 
Monte Carlo methods). 
This modeling depends also on the application. 
We discuss this point through the particular deconvolution problem in 
\ms. 

\subsection{Appropriate modeling of input signal}
We actually had started this discussion in previous section and we saw 
that, at least for linear inverse problems with a white Gaussian 
assumption of the noise, the posterior has for expression: 
$p(\xb|\yb)\propto \expf{-J(\xb)}$ 
with 
\beq \label{MAP_crit}
J(\xb)=\norm{\yb-\Hb\xb}^2 + \lambda\phi(\xb) 
\eeq
with $\phi(\xb)=-\ln p(\xb)$.  
Thus the expression and properties of $J(\xb)$, and consequently those of 
the posterior $p(\xb|\yb)$ depend on the prior $p(\xb)$. For example if 
$p(\xb)$ is Gaussian then 
\[
\phi(\xb)=-\ln p(\xb)=\sum_j x_j^2
\]
is a quadratic function of $\xb$. 
Then the MAP or PM estimates have the same values and their 
computation needs the optimization of a quadratic criterion which can be 
done either analytically or by using any simple gradient based algorithm. 
But the Gaussian modeling is not always an appropriate one. 
Let take our example of deconvolution of \ms data. We know \aprio that 
the input signal must be positive. Then a truncated Gaussian will be a 
better choice:  
\[
\phi(\xb)=\sum_j x_j^2, \hbox{~~~if~~} x_j\ge 0; 
\hbox{~~~else~~} \phi(\xb)=\infty.
\] 
But, we know still more about the input signal: it has 
pulse shapes, meaning that, if we look at the histogram of the samples 
of a typical signal, we see that great number of samples are near to zero 
but great deviations from this background are not rare. 
Thus, a generalized Gaussian 
\[
\phi(\xb)=\sum_j |x_j|^p \hbox{~~with~} 1\le p\le 2; 
\hbox{~~~if~~} x_j\ge 0; \hbox{~~~else~~} \phi(\xb)=\infty.
\] 
or 
a Gamma prior law 
\[
\phi(\xb)=\sum_j \ln x_j + x_j\; \hbox{if~} x_j\ge 0; 
\hbox{else~} \phi(\xb)=\infty.
\] 
would be better choices. 

We can also go further in details and want to account for the fact 
that we are looking for atomic pulses. Then we can imagine a binary 
valued random vector $\zb$ with $p(z_j=1)=\alpha$ and 
$p(z_j=0)=1-\alpha$, 
and describe the distribution of $\xb$ hierarchically: 
\beq
p(x_j|z_j) = z_j \, p_0(x_j)
\eeq
with $p_0(x_j)$ being either a Gaussian $p(x_j)=\Nc(m,\sigma^2)$ or 
a Gamma law $p(x_j)=\Gc(a,b)$. The second choice is more appropriate 
while the first results on simpler estimation algorithms. 
The inference can then be done through the joint posterior 
\beq 
p(\xb,\zb|\yb)\propto p(\yb|\xb)\, p(\xb|\zb) \, p(\zb)
\eeq
The estimation of $\zb$ is then called \emph{Detection} and that of $\xb$ 
\emph{Estimation}. 
The case where we assume 
$p(\zb)=\prod_j p(z_j)=\alpha^{n_1}(1-\alpha)^{(n-n_1)}$ 
with $n_1$ the number of ones and $n$ the length of the vector $\zb$,  
is called Bernoulli process and this modelization for $\xb$ is called 
\emph{Bernoulli-Gaussian} or 
\emph{Bernoulli-Gamma} as a function of the choice for $p_0(x_j)$. 

The difficult step in this modeling is the detection step which needs 
the computation of 
\beq 
p(\zb|\yb)\propto p(\zb) \intg p(\yb|\xb)\, p(\xb|\zb) \d{\xb}
\eeq 
and then its optimization over $\{0,1\}^n$ where $n$ is the length 
of the vector $\zb$. The cost of the computation of the exact solution 
is huge (a combinatorial problem). 

Many approximations to this optimization have been proposed which result 
to different algorithms for this detection-estimation problem 
\cite{Champagnat96a}. Many Monte Carlo techniques have also been proposed 
for generating samples of $\zb$ and $\xb$ from the posterior and thus 
compute the PM estimates of $\xb$. 
Giving more details on this modeling and details of corresponding 
algorithms is out of the scope of this paper. 

The results on the following figure illustrate this discussion. 
Here, we used the data in figure~1 and computed $\xb$ by optimizing 
the MAP criterion~(\ref{MAP_crit}), with different prior laws 
$p(\xb)\propto \expf{-\lambda\phi(\xb)}$ 
in between the following choices:\\ 
a) Gaussian: $\phi(\xb)=\sum x_j^2$, 
\\ 
b) Gaussian truncated on positive axis: $\phi(\xb)=\sum x_j^2, \; x_j>0$, 
\\ 
c) Generalized Gaussian truncated on positive axis:    
$\phi(\xb)=\sum |x_j|^p\;$ with $p=1.1, \; x_j>0 $. 
\\ 
d) Entropic prior $\phi(\xb)=\sum x_j \ln x_j - x_j, \; x_j>0$, 
\\ 
e) Gamma prior:  $\phi(\xb)=\sum \ln x_j +x_j, \; x_j>0$.

\clearpage\newpage
\bfig[hbt]
\btabu{c}
\includegraphics[width=\textwidth,height=4cm]{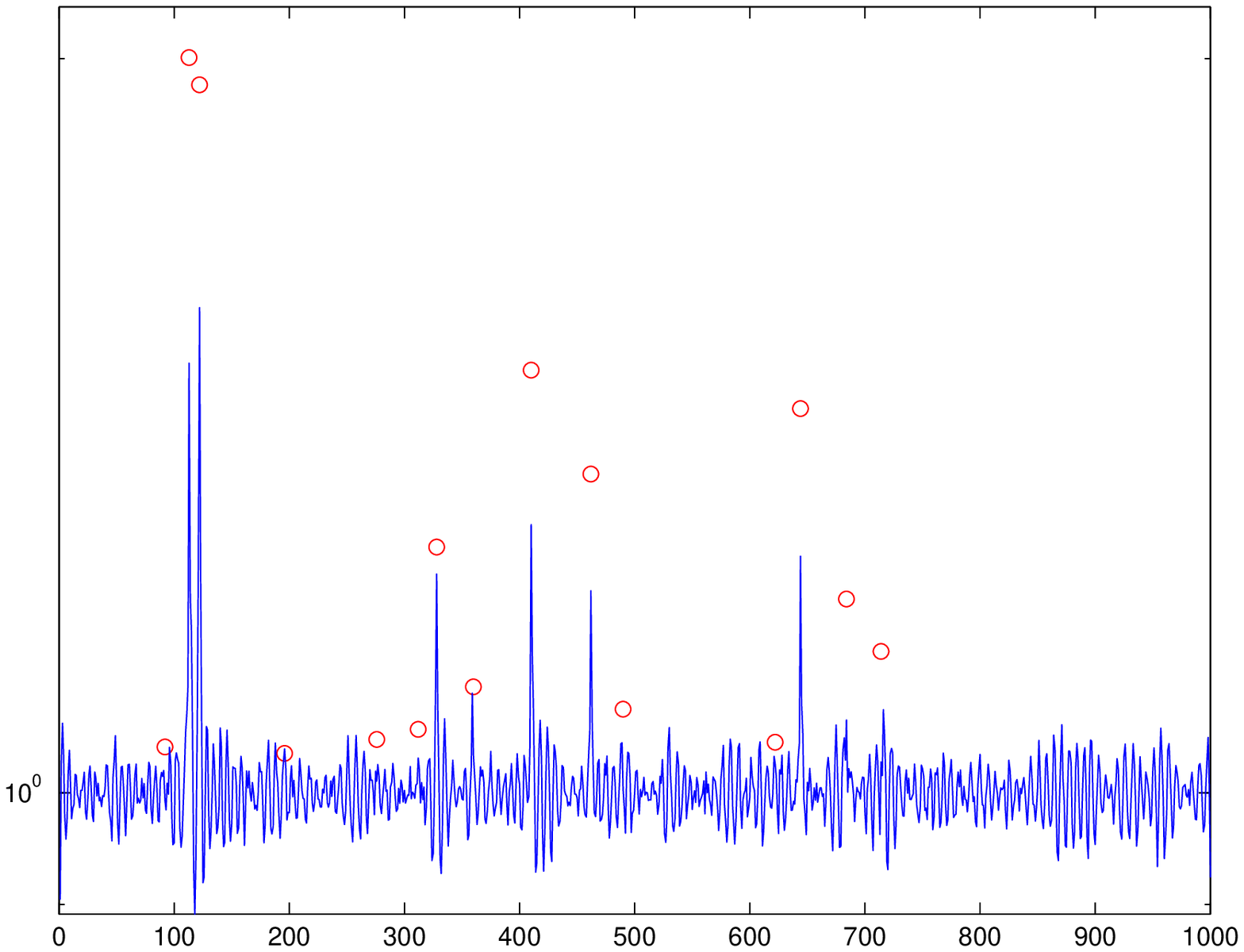}\\
\includegraphics[width=\textwidth,height=4cm]{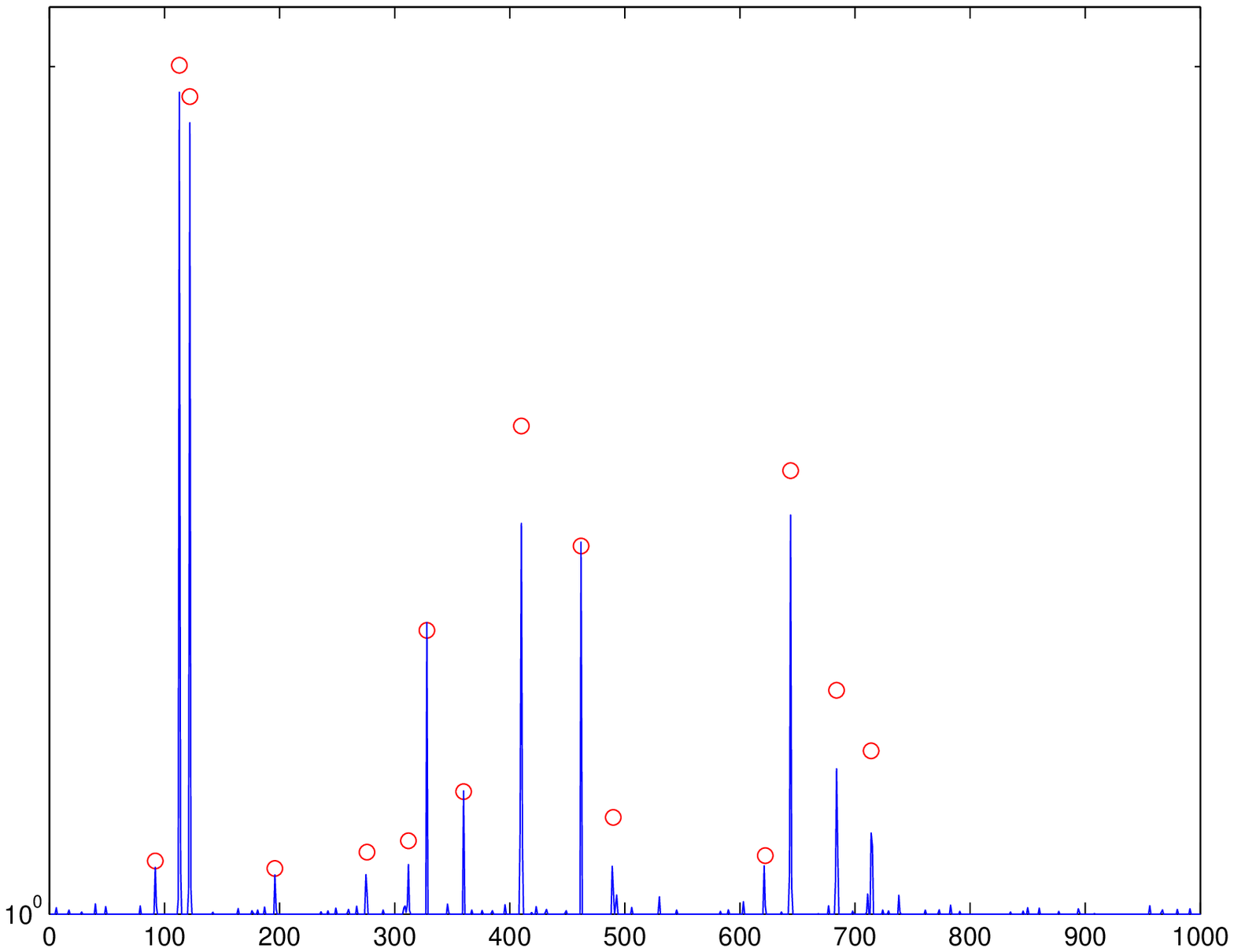}\\
\includegraphics[width=\textwidth,height=4cm]{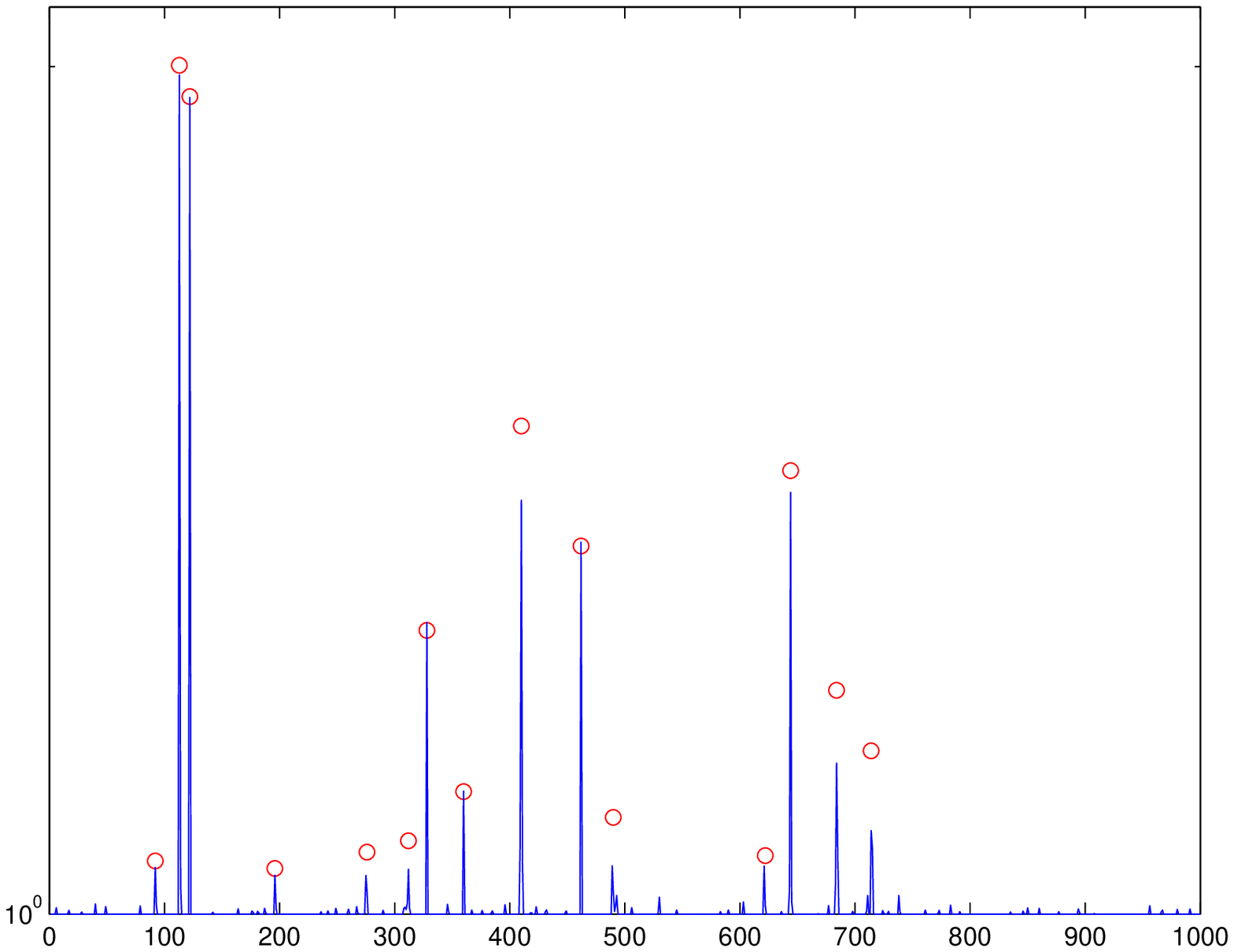}\\
\includegraphics[width=\textwidth,height=4cm]{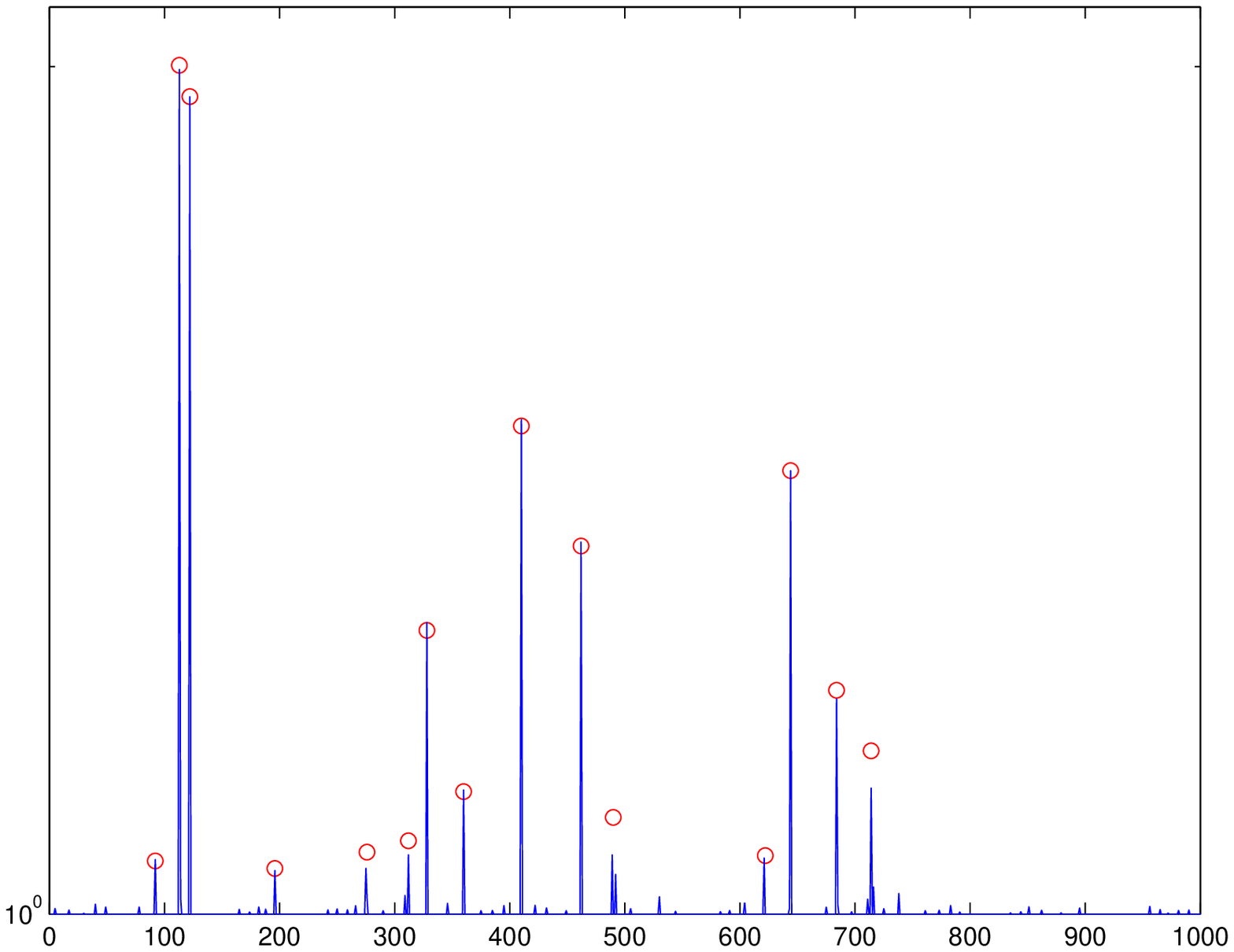}\\
\includegraphics[width=\textwidth,height=4cm]{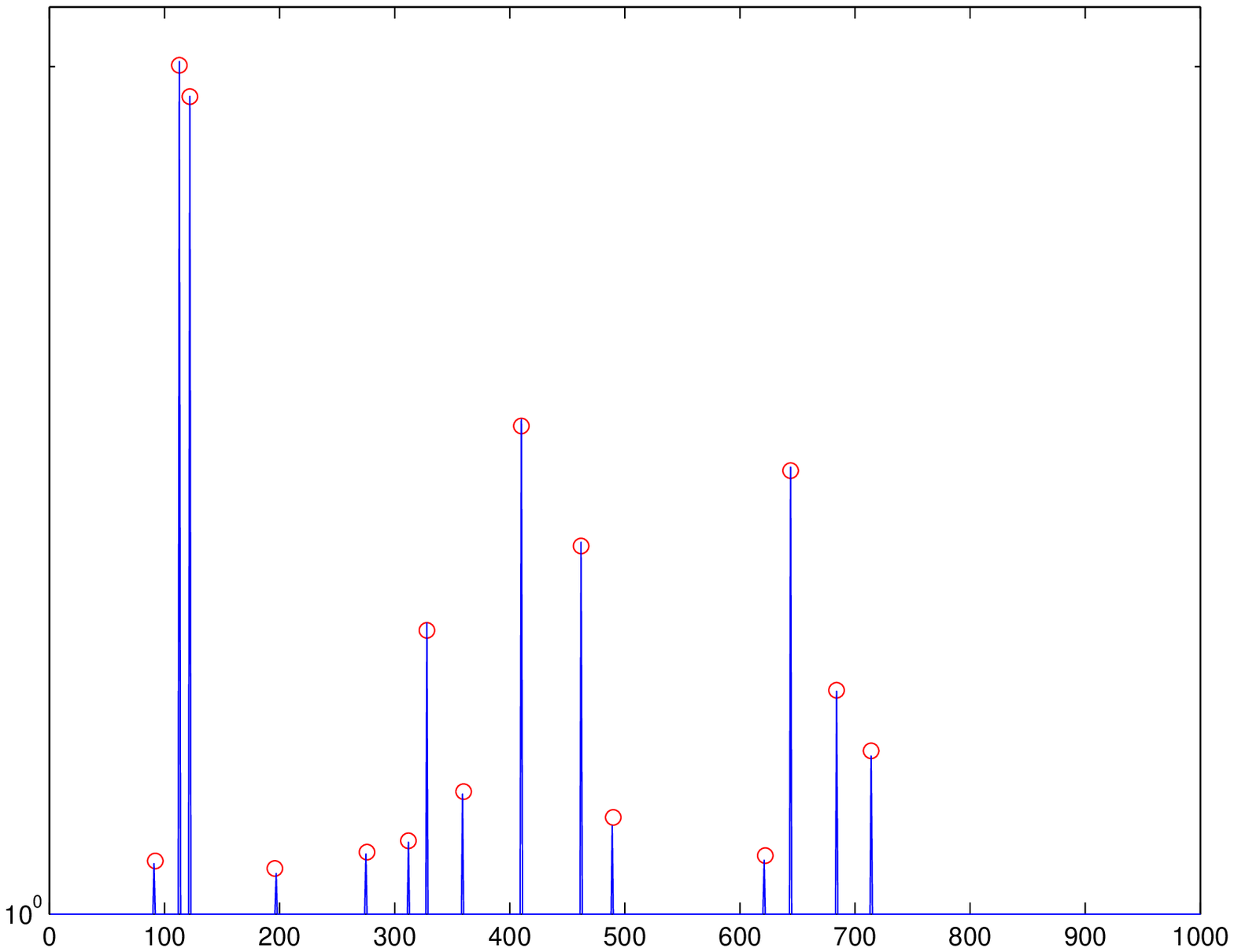}
\etabu
\caption
{Deconvolution results with different priors:\quad  
a) Gaussian\;   
b) Gaussian truncated to positive axis\; 
c) Generalized Gaussian. 
d) $-x\ln x$ entropic  prior \; 
e) $\ln x$ entropic or Gamma prior.  
 }
\label{fig4}
\efig

\clearpage\newpage
As it can be seen from these results
\footnote{Remark that the results are presented on a logarithmic scale for 
the amplitudes to show in more detail the low amplitude pulses. 
We used $\log(1+a)$ scale in place of $y$ scale which has the advantage 
of being equal to zero for a=0.}  
, for this application, 
the Gaussian prior does not give satisfactory result, but in 
almost all the other cases the results are more satisfactory, because the 
corresponding priors are more in agreement with the nature of the 
unknown input signal. 

\bfig[hbt]
\btabu{r} \includegraphics[width=\textwidth,height=10cm]{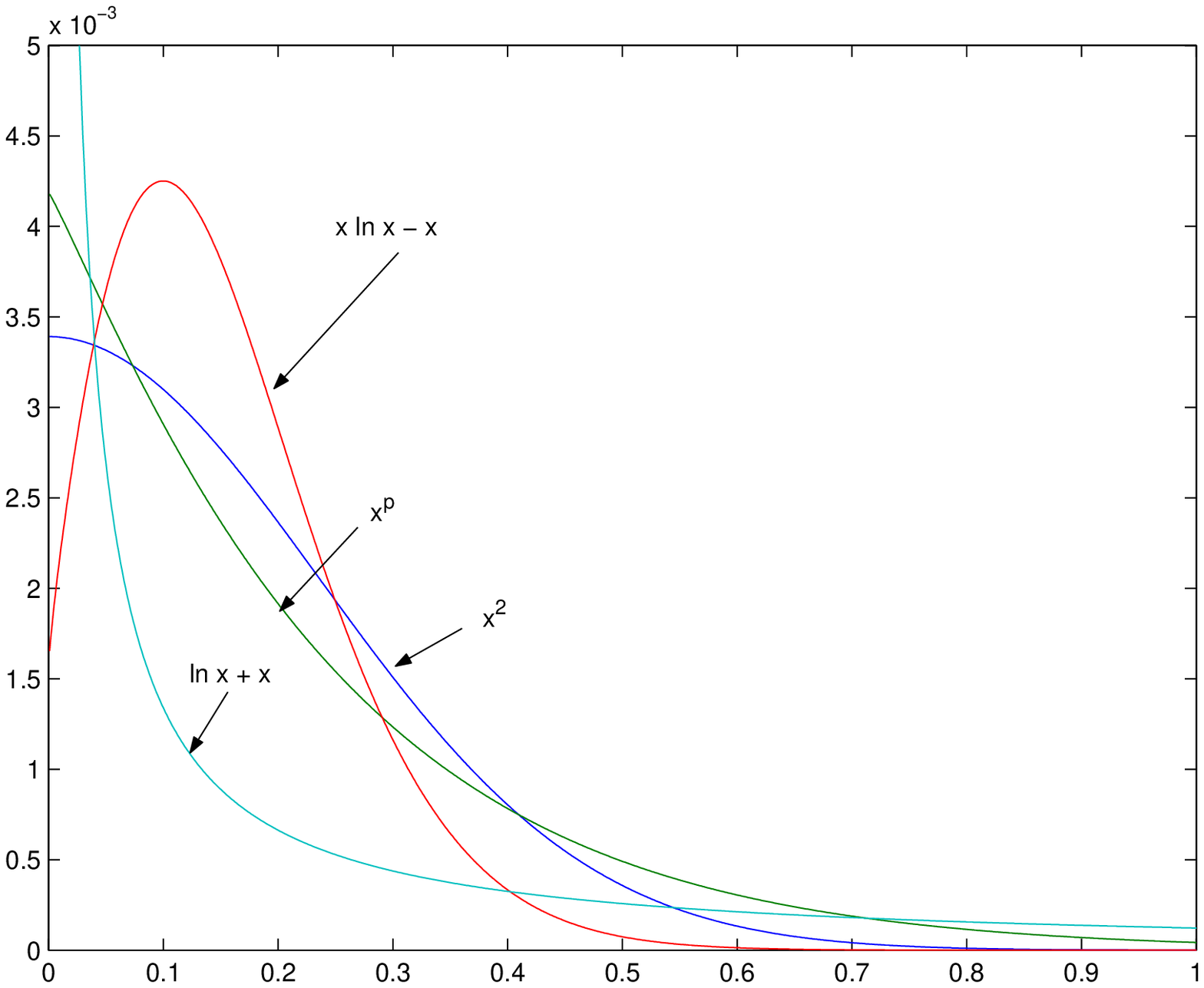} \etabu \\
\caption{Plots of the different prior laws 
$p(x)\propto\expf{-\lambda\phi(x)}$:
a) Truncated Gaussian \; $\phi(x)=x^2, \lambda=3$\;   
b) Truncated generalized Gaussian \; $\phi(x)=x^p,\;p=1.1,\;\lambda=4$; 
c) Entropic $\phi(x)=x\ln x - x, \lambda=10$\; 
d) Entropic $\phi(x)=\ln x + x, \lambda=0.1$.  
 }
\label{fig5}
\efig

The Gaussian prior (a) is not at all appropriate,  
Gaussian truncated to positive axis (b) is a better choice. 
The generalized Gaussian truncated to positive axis (c) 
and the $-x\ln x$ entropic priors (d) give also almost the 
same results than the truncated Gaussian case. 
The Gamma prior (e) seems to give slightly better 
result (less missing and less artifacts) than all the others. 
This can be explained if we compare the shape of all these priors 
shown in figure~(\ref{fig5}). 
The Gamma prior is sharper near 
to zero and has longer tail than other priors. It thus favorites 
signals with greater number of samples near to zero and still 
leaves the possibility to have very high amplitude pulses. 
However, we must be careful on this interpretation, because all 
these results depend also on the hyperparameter $\lambda$ 
whose value may be critical for this conclusion. In these experiments 
we used the same value for all cases. 
This brings us to the next open problem which is the determination 
of the hyperparameters. 

\subsection{Hyperparameter estimation}
\label{Hyperparameter} 
The Bayesian approach can be exactly applied when the direct 
(prior) probability laws $p(\yb|\xb,\phib_1)$ and $p(\xb|\phib_2)$  
are assigned. Even, when we have chosen appropriate laws, still 
we have to  determine their parameters $\phib=[\phib_1,\phib_2]$. 
This problem has been addressed by many authors 
and the subject is an active area in statistics. 
See \cite{Hall87,Hebert92,Johnson91,Titterington85},   
\cite{Younes88,Younes89,Bouman94,Fessler93,Liang92} 
and also \cite{Fortier93,Djafari93a,Djafari96b}.     

The Bayesian approach gives natural tools to handle this problem 
by considering $\phib=(\phib_1,\phib_2)$ as extra unknown parameters 
to infer on. We may then assign a prior law $p(\phib)$ to them 
too. However, the way to do this is also still an open problem. 
We do not discuss it more in this paper. The readers are invited to 
see~\cite{Kass94} for some extended discussions and references. 
When this step is done, we can again use the Bayesian approach 
and compute the joint posterior $p(\xb,\phib|\yb)$ from which 
we can follow three main directions:

\smallskip\noindent 
-- Joint MAP optimization: In this approach one tries to estimate 
both the hyperparameters and the unknown variables $\xb$ directly 
from the data by defining:
\beq
(\xbh,\phibh)=\argmax{(\xb,\phib)}{p(\xb,\phib|\yb)}   
\mbox{~where~}
p(\xb,\phib|\yb)
\propto p(\yb|\xb,\phib) \, p(\xb|\phib) \, p(\phib)   
\eeq
and where $p(\phib)$ is an appropriate prior law for $\phib$. 
Many authors used the non informative prior law for them.   
\\ 
-- Marginalization: The main idea in this approach is to distinguish 
between the two sets of unknowns: a high dimensional vector $\xb$ 
representing in general a physical quantity and 
a low dimensional vector $\phib$ representing the parameters of its 
prior probability laws. This argument leads to estimate 
first the hyperparameters by marginalizing over the unknown variables 
$\xb$: 
\beq \label{Hyper1}
p(\phib|\yb)\propto p(\phib) \, \intg 
p(\yb|\xb,\phib)\, p(\xb|\phib) \d{\xb}
\eeq
and then, using them in the estimation of the unknown 
variables $\xb$:  
\beq \label{Hyper2}
\phibh=\argmax{\phib}{p(\phib|\yb)} \lra  
\xbh =\argmax{\xb}{p(\xb|\yb,\phibh)}.
\eeq
Note also that when $p(\phib)$ is choosed to be uniform, then 
$p(\phib|\yb)\propto p(\yb|\phib)$ which is the likelihood of 
the hyperparameters $\phib$ and the corresponding maximum likelihood 
(ML) estimate has all the good asymptotic properties which may not 
be the case for the joint MAP estimation. However, for practical 
applications with finite data we may not care too much about the 
asymptotic properties of these estimates. 
\\ 
-- Nuisance parameters: In this approach the hyperparameters 
are considered as the nuisance parameters, so integrated out of 
$p(\xb,\phib|\yb)$ to obtain $p(\xb|\yb)$ and $\xb$ is estimated by 
\beq
\xbh =\argmax{\xb}{p(\xb|\yb)}\mbox{~~where~~}
p(\xb|\yb)=\intg p(\yb,\xb,\phib)\d{\phib}
\eeq
-- Joint Posterior Mean: Here, $\xb$ and $\phib$ 
are estimated as the posterior means:
\beq
\xbh=\esp{\xb|\yb}=\intg \xb \, p(\xb|\yb) \d{\xb} \mbox{~~and~~}  
\phibh=\esp{\phib|\yb}=\intg \phib \, p(\phib|\yb) \d{\phib}. 
\eeq
 
The main issue here is that, excepted the first approach, all the 
others need integrations for which, in general, there is not 
analytical expressions and their numerical computation cost 
may be very high. At the other hand, unfortunately, the estimation 
by the joint maximization has not the good 
asymptotic properties (when number of data goes to infinity) of the 
estimators obtained through the marginalization or expectation. 
However, in finite number of data, a comparison of their relative 
properties is still to be done. 
To see some more discussions and different possible implementations of 
these approaches see \cite{Djafari96b}. 
We have also to mention that, we can always use the 
Markov Chain Monte Carlo (MCMC) techniques 
to generate samples from the joint posterior $p(\xb,\phib|\yb)$ and then 
compute the joint posterior means and corresponding variances. It seems 
that these techniques are growing up. However, I see two main limitations 
for their application on real data: 
their huge computational cost and the need for some discussions on the 
tools to control their convergences. 

\subsection{Myopic or blind inversion problems}
\label{Blind_inversion}

Consider the deconvolution problems (\ref{1D_Convolution}) or 
(\ref{2D_Convolution}) and assume 
now that the psf $h(t)$ or $h(x,y)$ are partially known. 
For example, we know they have Gaussian shape, but the amplitude 
$a$ and the width $\sigma$ of the Gaussian are unknown. 
Noting by $\thetab=(a,\sigma)$ the problem then becomes the estimation 
of both $\xb$ and $\thetab$ from $\yb=\Hb_{\thetab}\xb+\epsilonb$. 
The case where we know only the support of the psf but not its shape 
can also be casted in the same way with $\thetab=[h(0),\ldots,h(p)]$ 

Before going more in details, we must note that, in general, 
the blind inversion problems are much harder than the simple 
inversion. Taking the deconvolution problem, we have seen in 
introduction that, the problem even when the psf is given is 
ill-posed. The blind deconvolution then is still more 
ill-posed, because here there are more fundamental under-determinations. 
For example, it is easy to see that, we can find an infinite number 
of pairs $(h, x)$ which result to the same convolution product 
$h*x$. This means that, to find satisfactory methods and algorithms 
for these problems need much more prior knowledge both on $x$ and on 
$h$, and in general, the inputs must have more structures (be rich 
in information content) to be able to obtain satisfactory results. 

Conceptually however, the problem is identical to the estimation of 
hyperparameters in previous section and any of the four approaches 
presented there can be used. 
One may wish however to distinguish 
between these parameters of the system $\thetab=(a,\sigma)$ and those 
hyperparameters of the prior law model descriptions 
$\phib=(\sigma_{\epsilon}^2,\sigma_{x}^2,\ldots)$. 
In that case, one can try to write down $p(\xb,\thetab,\phib|\yb)$ 
and use one of the following:

\smallskip\noindent 
-- Joint MAP estimation of $\xb$, $\thetab$ and $\phib$: 
\(\disp{
(\xbh,\thetabh,\phibh)
=\argmax{(\xb,\thetab,\phib)}{p(\xb,\thetab,\phib|\yb)}   
}\). 
\\
-- Marginalize over $\xb$ and estimate $\thetab$ and $\phib$ using: 
\(\disp{
(\thetabh,\phibh)=\argmax{(\thetab,\phib)}{p(\thetab,\phib|\yb)}  
}\)  
and then, estimate $\xb$ using:  
$\xbh=\argmax{\xb}{p(\xb|\yb,\thetabh,\phibh)}$.   
\\
-- Marginalize over $\xb$ and $\thetab$ and estimate $\phib$ using: 
\(\disp{
\phibh =\argmax{\phib}{p(\phib|\yb)}, 
}\),  
then estimate $\thetab$ using: 
$\thetabh =\argmax{\thetab}{p(\thetab|\yb, \phibh)}$ 
and finally, estimate $\xb$ using: 
$\xbh=\argmax{\xb}{p(\xb|\yb,\thetabh,\phibh)}$.   
\\
-- Joint Posterior Mean: Here, $\xb$, $\thetab$ and $\phib$ 
are estimated through their respective posterior means: 
$\xbh=\esp{\xb|\yb}$, $\thetabh=\esp{\thetab|\yb}$ and 
$\phibh  =\esp{\phib|\yb}$.
 
Here again, the joint optimization stays the simpler but we must be 
careful on interpretation of the results. For others, one can 
either use the Expectation-Maximization (EM) algorithms and/or MCMC 
sampling tools to approximately compute the 
necessary integration or expectation computations and overcome 
the computational cost issues. 

\section{Conclusions}
\label{Conclusions}

In this paper I presented a synthetic overview of methods for 
inversion problems starting by deterministic data matching and  
regularization methods followed by a general presentation of the 
probabilistic methods such as error probability law matching and 
likelihood based and the information theory and maximum entropy 
based methods. Then, I focused on the Bayesian inference. 
I show that, as it concerns the maximum \apost estimation 
method, one can see easily the link with regularization methods. 
We discussed however the superiority of the Bayesian framework 
which gives naturally the necessary tools for inferring the 
uncertainty of the computed solution, for the estimation of 
the hyperparameters or for handling myopic and blind inversion 
problems. We saw also that probabilistic modeling of signal and 
images is more flexible for introduction of practical prior 
knowledge about them. 
Finally, we illustrated some of these discussions through a 
deconvolution example in \ms data processing.  

\bibliographystyle{maxent95}	
\bibliography{bibfr,revuedef,baseAJ,baseKZ,gpipubli}
\end{document}